\documentclass[final,5p,times,twocolumn,authoryear]{elsarticle}

\usepackage{amssymb}
\usepackage{lipsum}

\usepackage[T1]{fontenc} 
\usepackage{stfloats}
\usepackage{graphicx}
\usepackage[normalem]{ulem}

\usepackage{txfonts}
\usepackage[table,xcdraw]{xcolor}
\usepackage{amssymb,amsmath,latexsym,mathrsfs}
\usepackage[hidelinks]{hyperref}

\journal{High Energy Astrophysics}

\begin{document}

\begin{frontmatter}



\title{Towards a new model-independent calibration of Gamma-Ray Bursts}

\author[a,b]{Arianna Favale}\ead{afavale@roma2.infn.it}
\author[c,d,e,f]{Maria Giovanna Dainotti}\ead{maria.dainotti@nao.ac.jp}
\author[b]{Adrià G\'omez-Valent}\ead{agomezvalent@icc.ub.edu}
\author[a]{Marina Migliaccio}\ead{migliaccio@roma2.infn.it}

\affiliation[a]{organization={Dipartimento di Fisica and INFN Sezione di Roma 2, Università di Roma Tor Vergata},
            addressline={via della Ricerca Scientifica 1}, 
            city={Roma},
            postcode={00133}, 
            country={Italy}}

\affiliation[b]{organization={Departament de Física Quàntica i Astrofísica and Institut de Ciències del Cosmos, Universitat de Barcelona},
            addressline={Av. Diagonal 647}, 
            city={Barcelona},
            postcode={08028}, 
            country={Spain}}

\affiliation[c]{organization={Division of Science, National Astronomical Observatory of Japan}, addressline={2-21-1 Osawa, Mitaka}, city={Tokyo}, postcode={181-8588}, country={Japan}}

\affiliation[d]{organization={The Graduate University for Advanced Studies (SOKENDAI)}, addressline={Shonankokusaimura, Hayama, Miura District}, city={Kanagawa}, postcode={240-0115}, country={Japan}}

\affiliation[e]{organization={Space Science Institute}, addressline={4765 Walnut St Ste B}, city={Boulder}, CO postcode={80301}, country={USA}}

\affiliation[f]{organization={Nevada Center for Astrophysics, University of Nevada}, addressline={4505 Maryland Parkway}, city={Las Vegas}, NV postcode={89154}, country={USA}}

\begin{abstract}
Current data on baryon acoustic oscillations and Supernovae of Type Ia (SNIa) cover up to $z\sim 2.5$. These low-redshift observations play a very important role in the determination of cosmological parameters and have been widely used to constrain the $\Lambda$CDM and models beyond the standard, such as the ones with open curvature. To extend this investigation to higher redshifts, Gamma-Ray Bursts (GRBs) stand out as one of the most promising observables. In spite of being transient, they are extremely energetic and can be used to probe the universe up to $z\sim9.4$. They exhibit characteristics that suggest they are potentially standardizable candles and this allows their use to extend the distance ladder beyond SNIa. The use of GRB correlations is still a challenge due to the spread in their intrinsic properties. One of the correlations that can be employed for the standardization is the fundamental plane relation between the peak prompt luminosity, the rest-frame end time of the plateau phase, and its corresponding luminosity, also known as the \textit{three-dimensional Dainotti correlation}. In this work, we propose an innovative method of calibration of the Dainotti relation which is independent of cosmology. We employ state-of-the-art data on Cosmic Chronometers (CCH) at $z\lesssim2$ and use the Gaussian Processes Bayesian reconstruction tool.  To match the CCH redshift range, we select 20 long GRBs in the range $0.553 \leq z \leq 1.96$ from the \textit{Platinum sample}, which consists of well-defined GRB plateau properties that obey the fundamental plane relation. To ensure the generality of our method, we verify that the choice of priors on the parameters of the Dainotti relation and the modelling of CCH uncertainties and covariance have negligible impact on our results. Moreover, we consider the case in which the redshift evolution of the physical features of the plane is accounted for. We find that the use of CCH allows us to identify a sub-sample of GRBs that adhere even more closely to the fundamental plane relation, with an intrinsic scatter of $\sigma_{int}=0.20^{+0.03}_{-0.05}$ obtained in this analysis when evolutionary effects are considered. In an epoch in which we strive to reduce uncertainties on the variables of the GRB correlations in order to tighten constraints on cosmological parameters, we have found a novel model-independent approach to pinpoint a sub-sample that can thus represent a valuable set of standardizable candles. This allows us to extend the cosmic distance ladder presenting a new catalogue of calibrated luminosity distances up to $z=5$.
\end{abstract}


\end{frontmatter}




\section{Introduction}
Standard candles constitute a well-established tool to measure cosmic distances in our universe, which in turn are used to constrain cosmological parameters of fundamental importance. 
They played a pivotal role in the discovery of the late-time cosmic acceleration \citep{SupernovaSearchTeam:1998fmf,SupernovaCosmologyProject:1998vns} and in the last years
have proved also relevant in the discussion of the Hubble tension \citep{Riess:2021jrx}.
Nowadays, the mismatch between Hubble parameter values estimated independently by late- and early-time probes through the direct and inverse distance ladders has reached the $\approx 4\sigma-6 \sigma$ level depending on the particular data set used (see \cite{Verde:2019ivm,di2021realm,Dainotti_2021,Perivolaropoulos:2021jda,Abdalla:2022yfr,Dainotti2022Galax..10...24D, Riess:2022oxy,Verde:2023lmm} for dedicated reviews). There exists a large redshift gap between measurements of Baryon Acoustic Oscillations (BAO) and Supernovae of Type Ia (SNIa), which cover up to $z\sim 2.5$, and CMB observations, which also encode information from the last-scattering surface ($z\sim1100$) and beyond. Data in that gap could enhance our understanding about the underlying physical mechanism responsible for the expansion of the universe. They might be crucial to understand the origin of the cosmological tensions (see e.g. \cite{Gomez-Valent:2023uof}), including the anomalous existence of very massive galaxies in the early universe \citep{2023Natur.616..266L}. Moreover, recently a mismatch has been found between $\Lambda$CDM predictions and the Hubble diagram built out from high-$z$ quasars (QSOs) \citep{Risaliti:2018reu,lusso2019,Bargiacchi2021, signorini2023building}, which have been employed as standardizable candles \citep{Risaliti:2015zla,Risaliti:2016nqt,lusso2020quasars,DainottiQSO,Colgain:2022nlb,LenartbiasfreeQSO2022,Dainotti2023alternative,bargiacchi2023gamma}. Therefore, it has now become necessary to extend the research to new cosmological probes at intermediate redshifts, which can come to our aid, widening the unexplored range of redshifts and serving as distance indicators. Exploring higher $z$ requires observing more luminous objects than SNIa. This can be the case not only of QSOs, which reach up to $z\sim7.5$ \citep{banados2018}, but also of Gamma-Ray Bursts (GRBs), the most intense explosions in the universe after the Big Bang. These objects have been detected up to $z\sim9.4$ so far \citep{Cucchiara2011}, but their observation might reach $z\sim20$ \citep{Lamb2003}. The great advantage of the use of GRBs is their coverage at much higher redshifts than SNIa and BAO. Hence, it is important to improve their calibration in the range where other sources are present.

To standardize GRBs and use them as cosmological tools, it is necessary to find tight and intrinsic relationships, not induced by selection biases, between parameters of their light curves (or spectra) and their luminosity (or energy).
In particular, in this work, we will focus on the relation involving the features of the plateau emission phase, that is, the luminosity at the end of the plateau emission and its rest frame duration \citep{dainotti2008time,Dainotti2010,dainotti11a,Dainotti2013,dainotti17a,Dainotti2020b,Dainotti2022c}. This two-dimensional (2D) relation has been used as a cosmological tool for 15 years \citep{cardone2009updated, cardone2010constraining, Dainotti2013b}. However, considering three or more parameters could lead to tighter correlations. One of these correlations is the so-called \textit{three-dimensional (3D) Dainotti relation}. It relates the peak prompt luminosity, the rest-frame end time of the plateau, and its corresponding luminosity \citep{dainotti2008time, Dainotti_2016, Dainotti_2017}. 
The main advantages of using the Dainotti relation are that it is supported by a fundamental physical process, the magnetar emission \citep{rowlinson2014constraining, bernardini2015gamma, rea2015constraining, stratta2018magnetar, DallOsso:2023gdk}, and it overcomes selection biases that could invalidate the reliability of the relations themselves \citep{Dainotti2013,dainotti17a}. Moreover, prompt-afterglow relations present a reduced scatter in the afterglow features compared to more varied prompt emission properties \citep{Dainotti22}. This is yet another reason why using the plateau emission for standardizing GRBs might be more recommended.
There are many correlations that help broaden our understanding of GRBs and their possible application as a cosmological tool. For instance, the $E_{p}-E_{iso}$ Amati correlation connects the spectral peak energy in the GRB cosmological rest frame and the isotropic equivalent radiated energy \citep{amati2002intrinsic, amati2008measuring}. \cite{Yonetoku:2003gi} correlated $E_{p}$ instead with the peak luminosity, $L_{p}$; this is called in the literature the Yonetoku relation. \cite{Norris2000} has also established a relation between $L_{p}$ and the spectrum lag, the so-called $L_{p} - \tau_{lag}$ relation, and the bulk Lorentz factor in the context of the GRB fireball scenario is also found to correlate with $E_{iso}$ \citep{liang2015} or with $L_{iso}$ \citep{ghirlanda2012}.
However, the issue with many of these correlations is that they have not been demonstrated to be reliable after the truncation and selection effect tests, and when these corrections are applied, the dispersion of the relation itself increases \citep{Collazzi2011, Heussaff:2013sva, Petrosian:2015vda}. The source of this scatter is due to both instrumental biases and the underlying mechanism that governs the GRB emission, such as the energy and frequency at which it occurs. Indeed, the physical meaning of many of these relations is still under debate after many years since more than one interpretation is often viable, showing that the emission processes that rule GRBs still need further investigations. All these aspects might significantly limit the applicability of such relations for cosmological purposes. We refer the reader to \cite{Dainotti_Amati17} for a discussion on selection and instrumental biases and to \cite{2018AdAst2018E...1D} or \cite{2022Univ....8..310P} for a comprehensive review of the several correlations existing in the literature and their possible physical explanation.

So far, GRBs data and their correlations have been used for many applications in cosmology, such as the investigation of the components of our universe, the nature of dark energy, and the Hubble constant tension (see, e.g., \cite{Amati2019,Khadka2020,Khadka:2021vqa, cao_gamma-ray_2022}). However, if a cosmological model is assumed when calibrating GRBs (or other standardizable objects, like SNIa), the so-called \textit{circularity problem} \citep{Ghisellini_Ghirlanda_Firmani_Lazzati_Avila-Reese_2005,ghirlanda2006gamma, wang2015gamma} can be encountered if the resulting correlations are then used to constrain parameters of models different from the one employed in the calibration. Up to now, two solutions have been proposed to overcome this problem. One is to fit simultaneously the correlation parameters and the parameters of a cosmological model of interest using GRB data \citep{ghirlanda2004gamma, li2008overcoming, amati2008measuring, postnikov14,cao_gamma-ray_2022, cao2022standardizing,dainotti2023gamma}. In this way, cosmological constraints are uniquely determined by GRBs. 

On the other hand, based on the idea of the distance ladder, one can calibrate GRBs with low-redshift probes, such as calibrated SNIa \citep{liang2008cosmology,postnikov14}, given that objects at the same redshift should have the same luminosity distance regardless of the underlying cosmology if the latter respects the cosmological principle (CP). The distance ladder measurement is basically model-independent since it only relies on the CP and the assumption that SNIa are good standardizable objects, i.e., with a standardized absolute magnitude that remains constant from our vicinity to the far end of the cosmic ladder. One uses other standard candles, such as Cepheids \citep{riess2021uncrowding} or the Tip of the Red Giant Branch \citep{freedman2020calibration, Freedman_2021, Freedman:2023jcz} in the lower rungs of the ladder to calibrate the SNIa, and then uses SNIa to calibrate GRBs in a model-independent way. Most of the calibrated-GRB cosmological analyses are based on this idea. They make use of cosmographical methods \citep{luongo2020kinematic, Mu:2023bsf} or Gaussian Processes (GP) as in \cite{liang_calibrating_2022}. The calibration occurs at $z\lesssim 2.5$ (i.e. below the maximum redshift in the SNIa samples) and then the calibrated GRB relations can be employed to build an extended Hubble diagram up to the high-$z$ region covered by the GRBs. In doing so, analogously to the calibration of SNIa, it is usually assumed that the GRB correlation does not evolve with redshift, although this evolution could actually have a non-negligible impact \citep{kumar_gamma_2023}. 
For a discussion on how different is the calibration of GRBs on SNIa or independent from SNIa and how the evolution impacts the outcome see \cite{Dainotti2013,Dainotti2017NewAR..77...23D,dainotti2023gamma}.
Another possible caveat of this method is that some unaccounted-for systematic biases of SNIa may propagate into the calibration results. Avoiding these biases might be important in light of the Hubble tension. 

Looking for alternative calibrators becomes thus fundamental. In the future, the use of standard sirens up to $z\sim4$ appears to be a promising route \citep{wang2019calibration}.
As of now, data on galaxy clusters in the redshift range $0.14 \leq z \leq 0.89$ have, for instance, already been employed in the calibration of GRBs \citep{govindaraj2022low}. Moreover, \cite{Amati2019} recently introduced the use of Observational Hubble Data (OHD) to calibrate the Amati relation, something that has been further exploited using GP \citep{li_testing_2023, kumar_gamma_2023} and other techniques, such as the Bézier parametric curve obtained through the linear combination of Bernstein basis polynomials \citep{Montiel, luongo2020kinematic, luongo2021model, luongo_intermediate_2022}. All these analyses aim to build an extended Hubble diagram through the calibration of GRBs, and then combine them with other data sets (e.g., SNIa, BAO) to study and constrain different cosmological models.
There are several works related to the application of the 3D Dainotti relation as a cosmological tool together with other probes \citep{dainotti2023gamma,Dainotti2023alternative,bargiacchi2023gamma}. In the last two papers, a discussion on the most appropriate likelihood assumptions must be made since while the Gaussian is the most appropriate assumption for GRBs it is not for the other probes including SNIa \citep{Dainotti2024JHEAp..41...30D}.

In this work, we aim to present a novel model-agnostic method to understand whether low-redshift calibrators can univocally determine the correlation parameters of GRBs, in order to use them as distance indicators. It is even more powerful than previous methods.  In particular, we focus on the role played by the data on the Hubble function obtained from Cosmic Chronometers (CCH) \citep{Jimenez:2001gg} in calibrating the Dainotti correlations using a specific data set of GRBs, called the Platinum sample.
Recently, the 2D Dainotti relation has been calibrated with GP and OHD by \cite{Hu2021} and \cite{Wang_2022}, whose approach has been extended by \cite{Tian:2023ffe} also for the 3D relation. However, these authors adopt different OHD and GRB data sets from those considered by us in this work. \cite{Tian:2023ffe}, in particular, use a combination of 31 $H(z)$ measurements from CCH and 5 from BAO data, after calibrating the latter assuming standard pre-recombination physics. The GRBs that they analyze are collected according to a different sample selection (radio plateau phases instead of X-ray plateau phases) and at different redshifts with respect to the Platinum sample. In addition, they do not take into account correlations between different CCH data points and, for the fitting analysis, they use a slightly different likelihood based on the D'Agostini method \citep{dagostini2005}.
In our work, we employ the GP-reconstruction of the luminosity distance to calibrate the Dainotti relations, varying all the quantities that define the plane. We test the robustness of our results by selecting different priors on the parameters of the Dainotti relations and studying the impact of the covariance of the reconstructed luminosity distance. We do not use data on $H(z)$ from BAO to avoid the need to make model-dependent assumptions about the physics at the decoupling time in the calibration process. Moreover, we provide and justify a suitable set-up for the GP training and extend our methodology to account for the redshift-evolution corrections of the Dainotti relation, showing how this is actually of fundamental importance for the extension of the distance ladder.

This paper is organized as follows: in Sec. \ref{sec:dainotticorr}, we present and describe the two- and three-dimensional Dainotti relations, the latter defines the fundamental plane of GRBs; in Secs. \ref{sec:data} and \ref{sec:method}, we describe the data sets and methodology employed in this work, respectively; in Sec. \ref{sec:results} we present our results, discussed in detail in Sec. \ref{sec:discussion}. Finally, we draw our conclusions in Sec. \ref{sec:conclusion}.

\section{The Dainotti Correlations}\label{sec:dainotticorr}

The three-dimensional fundamental plane correlation, also known as 3D Dainotti correlation, reads \citep{Dainotti_2016,Dainotti_2017,dainotti2020x}

\begin{equation}\label{eq:3DDainotti}
    \log L_{\rm X} = C_o + a \log T^*_{\rm X} + b \log L_{\rm peak}\,,
\end{equation}%
where
\begin{equation}\label{eq:Lx}
    L_{\rm X} = 4\pi D^2_L F_{\rm X} K_{\rm plateau}\,
\end{equation}%
is the X-ray source rest-frame luminosity,
\begin{equation}\label{eq:lpeak}
    L_{\rm peak} = 4\pi D^2_L F_{\rm peak} K_{\rm prompt}
\end{equation}%
is the peak prompt luminosity (both luminosities are in units of $erg \ s^{-1}$) and $T^*_{\rm X}$ is the characteristic time scale which marks the end of the plateau emission (in units of $s$) \footnote{All the quantities appearing in the logs of Eq. \eqref{eq:3DDainotti} are made dimensionless, dividing them by their corresponding units.}. In the above equations, $F_{\rm X}$ is the measured X-ray energy flux at $T^*_{\rm X}$, $F_{\rm peak}$ is the measured gamma-ray energy flux at the peak of the prompt emission over a 1$s$ interval (both fluxes are in units of $erg \ cm^{-2} \ s^{-1}$), $K_{\rm plateau}=(1+z)^{\alpha_{\rm plateau}-2}$ is the power-law plateau $K$-correction with $\alpha_{\rm plateau}$ being the X-ray photon index of the plateau phase. The $K$-correction for the prompt emission, $K_{\rm prompt}$, depends instead on the X-ray photon index ($\alpha_{\rm prompt}$), energy and differential spectrum. 
Only the GRBs for which the spectrum calculated at 1 second has a smaller $\chi^{2}$ for a single power-law (PL) fit than for a cut-off power-law (CPL) are considered. In particular, following \cite{Sakamoto2011}, when $\Delta \chi^{2} = \chi^{2}_{\rm PL} - \chi^{2}_{\rm CPL} >6$, the CPL model is preferred over the PL to compute $K_{\rm prompt}$ (see also \cite{Dainotti_2016} for details).

The correlation parameters in Eq. \eqref{eq:3DDainotti} are identified by $a, b$, and $C_o$, where the $a$ parameter denotes the anti-correlation between $\log L_X$ and $\log T^{*}_{\rm X}$ possibly driven by the magnetar \citep{rowlinson2014constraining,rea2015constraining}, $b$ the relation between $\log L_{\rm peak}$ and $\log L_X$, and $C_o$ is the normalization of the plane. 
When dealing with GRBs, one has also to consider that these data suffer from an unknown source of scatter on the plane, $\sigma_{int}$, which cannot be neglected since it specifies the tightness of the relation. 
This tightness relies on the spin period and magnetic field variations of a fast millisecond magnetar. Indeed, the plateau emission can be ascribed to the magnetar model \citep{dall2010grb,rowlinson2014constraining, rea2015constraining, stratta2018magnetar}, which describes that the X-ray plateaus are caused by a fast-spinning neutron star. In this case, the slope of the correlation, which is identified by $a$, is equal to -1. Other physically motivated interpretations have been discussed. For instance, in \cite{vanEerten:2014gka} it is shown that both $L_{\rm peak}-L_X$ and $L_X-T^{\star}_X$ correlation can be recovered within the standard forward shock model for GRB afterglows, with the addition of the microphysical parameters. We will see that our results are fully compatible with the magnetar model predictions.

While $T^*_{\rm X}, F_{\rm X}, \alpha_{\rm plateau}, F_{\rm peak}, \alpha_{\rm prompt}$ are measurable quantities, $a, b, C_o$, and $\sigma_{int}$ can only be determined through a proper calibration, which requires the use of cosmological luminosity distances. It is immediate to notice in Eqs. \eqref{eq:Lx} and \eqref{eq:lpeak} that both $L_{\rm X}$ and $L_{\rm peak}$ depend on the background cosmology since the luminosity distance $D_L$ requires the knowledge of the functional form of the Hubble expansion rate. Assuming a flat Universe\footnote{The TT,TE,EE CMB likelihood from {\it Planck} prefers a closed universe at the $\gtrsim 2\sigma$ level in the context of $\Lambda$CDM \citep{Aghanim:2018eyx,Handley:2019tkm,DiValentino:2019qzk}. However, when data on BAO, SNIa, the full-shape galaxy power spectrum or CCH are also considered in the fitting analysis, the compatibility with spatial flatness (i.e., with $\Omega_k=0$) is recovered \citep{Aghanim:2018eyx,Efstathiou:2020wem,Vagnozzi:2020rcz,Vagnozzi:2020dfn,deCruzPerez:2022hfr,deCruzPerez:2024shj}. Model-independent studies with low-$z$ data reach also this conclusion, but with a larger uncertainty \citep{Collett:2019hrr, Dhawan:2021mel, Favale:2023lnp, Qi:2023oxv}. See \cite{DiValentino:2020srs} for a review.},

\begin{equation}\label{eq:DL}
    D_L(z)= c(1+z) \int^z_0 \frac{dz'}{H(z')} \,,
\end{equation}%
where the theoretical formulation of $H(z)$ is defined in Eq. \eqref{eq:hz_cch}.
In this work, we leverage the availability of data on $H(z)$ from cosmic chronometers, which do not rely on very strong cosmological assumptions. We propose to use these measurements together with GP to reconstruct the shape of $H(z)$ and obtain agnostic estimates of the luminosity distance at the redshifts of the GRB data, $z_{\rm GRB}$.

As we will see later on, the model-independent analysis performed with the 3D Dainotti relation only sets an upper bound on the value of $b$, being our constraint on this parameter compatible with $b=0$ at already $68\%$ C.L. This is possibly pointing out that the contribution of the peak luminosity in the prompt emission is not needed for the use of the correlation as a viable cosmological tool, which reduces therefore to an X-ray time-luminosity relation, i.e. the \textit{2D Dainotti correlation} \citep{dainotti2008time}. Neglecting $b$ in Eq. \eqref{eq:3DDainotti}, this relation is then defined as
\begin{equation}\label{eq:2DDainotti}
    \log L_{\rm X} = C_o + a \log T^*_{\rm X}\,.
\end{equation}%
The 2D Dainotti relation rules out basic thin shell models but not the basic thick shell model \citep{vanEerten:2014gka}. In the latter case, both forward- and reverse shock-dominated outflows are shown to be consistent with this relation. When a slope of -1.2 is observed, it means that accretion onto the black hole is favored, as discussed in \cite{cannizzo2009new, cannizzo2011fall}.

We will apply our methodology to the calibration of both, the 3D and 2D Dainotti relations.

\section{Data}\label{sec:data}

In this section, we describe the data sets employed in this work. 

\subsection{The Platinum sample}\label{sec:platinum}

Observations of 50 long GRBs (i.e., with a burst phase that lasts $>2s$) in the redshift range $0.553 \leq z \leq 5.0$ constitute the so-called \textit{Platinum sample}, defined for the first time in \cite{dainotti2020x}. It is a sub-sample of the larger Gold Sample \citep{Dainotti_2016}. The Platinum sample collects well-defined morphological features of GRBs that show plateaus with an inclination $< 41^{\circ}$ and that last $>500 s$ and with no flares. In terms of measured quantities at a given redshift $z$, we find five properties: $T^*_{\rm X}, F_{\rm X}, K_{\rm plateau}, F_{\rm peak}, K_{\rm prompt}$. These quantities can be related together to form a fundamental plane for the GRBs which finds its expression in the 3D Dainotti correlation in Eq. \eqref{eq:3DDainotti}, or the 2D Dainotti correlation in Eq. \eqref{eq:2DDainotti} if one removes the contribution of the peak prompt luminosity. For details about the physical meaning of the GRBs properties and the correlations, we refer the reader to Sec. \ref{sec:dainotticorr}.

\begin{figure}[t!]
    \centering
    \includegraphics[width=\columnwidth]{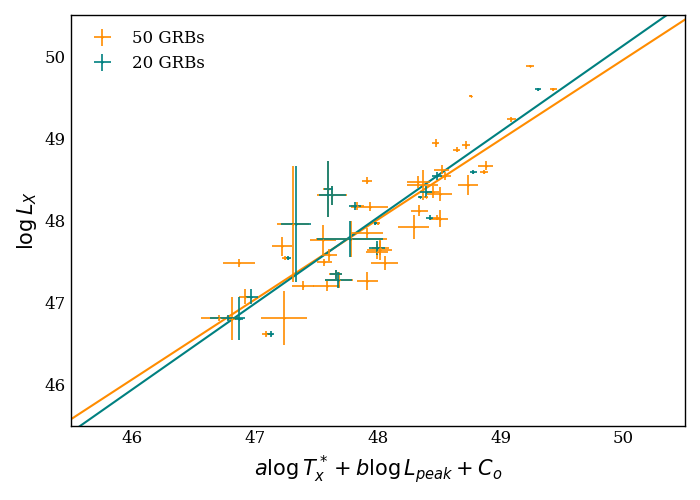}
    \caption{2D projections of the fundamental plane for the Platinum sample (in orange) and for the sub-sample of 20 GRBs used in this work (in green), both obtained taking the mean values of $a,b$, and $C_o$ from the related fundamental plane fitting analyses. The error bars represent the 1$\sigma$ uncertainty.}
    \label{fig:FP_bf}
\end{figure}

In Fig. \ref{fig:FP_bf}, we present the 2D projection of the fundamental plane for both the full Platinum sample (50 GRBs) and the sub-set of 20 GRBs that fall within the redshift range covered by the CCH data, which are used in our calibration of the Dainotti relations (see Secs. \ref{sec:cch} and \ref{sec:method}, and Tables \ref{tab:20platinum} and \ref{tab:CCH}). Here we fix the parameters $a,b$, and $C_o$ that enter the 3D relation to the values obtained from the fundamental plane fitting analysis. This is the standard procedure, widely used in the context of GRB cosmology. The likelihood is built by making use of the D'Agostini and Kelly method \citep{dagostini2005,Kelly2007}, and an underlying cosmological model is assumed to compute the luminosity distances entering Eqs. \eqref{eq:Lx} and \eqref{eq:lpeak}. An example of its application can be found in \cite{Dainotti22}, where a flat $\Lambda$CDM model with $H_0 = 70$ km/s/Mpc is considered, while $\Omega_M$ is free to vary. We follow the same approach as in that reference in Fig. \ref{fig:FP_bf}, just to visualize where the GRBs that constitute the Platinum sample are located in the plane according to the 3D Dainotti relation. The results obtained with 50 GRBs read $a=-0.88\pm0.12, b=0.54\pm0.12$, $C_o=23.04\pm6.23$ and $\sigma_{int} = 0.36\pm0.04$. For the results obtained with 20 GRBs, we refer instead the reader to the constraints listed in Table \ref{tab:prior_3D}. We will use them later in the analysis to test the impact of the prior choice on the parameters themselves. As a last note, we add here that in order to check whether the sub-sample of 20 GRBs is representative of the full Platinum sample, we perform a Kolmogorov-Smirnov (KS) test \citep{Kolmogorov,Smirnov} for all the variables of the Platinum sample used in this work, i.e. $\log T^{\star}_X$, $F_{\rm peak}$, $K_{\rm prompt}$, $\log F_X$ and $K_{\rm plateau}$. We also cross-check the results with the Anderson-Darling (AD) test \citep{ADtest}. These statistical analyses allow us to estimate the significance level for rejecting or 
 not rejecting the null hypothesis that the two samples come from the same parent distribution. The results of both tests show that we cannot reject the null hypothesis at 95\% C.L., as we find a $p$ value greater than 0.05 in all the tested variables. This supports the choice of the GRBs sub-set used in the calibration analysis and this does not introduce any bias in this choice as no statistical difference in the sample is shown. We present the results of KS and AD tests in \ref{app:KSAD}.

\subsection{Cosmic Chronometers}\label{sec:cch}
With the minimal assumption that the Universe is described by a Friedmann-Lemaître-Robertson-Walker metric, where the scale factor $a(t)$ is directly related to the redshift as $z = a^{-1} -1$, the expansion rate of the universe can be written as follows
\begin{equation}\label{eq:hz_cch}
    H(z)= \frac{\dot{a}}{a} = -\frac{1}{1+z}\frac{dz}{dt}\,,
\end{equation}
that is, the Hubble function is related to the differential ageing of the universe $dt$ as a function of $z$. However, the ratio $dz/dt$ is not directly observable and one needs to identify objects in the Universe whose time evolution across redshift ranges can be accurately determined, i.e. \textit{cosmic chronometers}. This can be done with the help of differential age techniques \citep{Jimenez:2001gg}. The best candidates for CCH are massive, passively evolving galaxies formed at $z\sim 2-3$ over a brief period, typically $t\sim0.3$ Gyr. These galaxies are homogeneous over cosmic time, meaning that they have formed at the same time independently of the redshift at which they are observed. Indeed, they evolve on timescales much longer than their differential ages. This allows us to use the difference in their evolutionary states to reveal the time elapsed between the redshifts, i.e. how much time has passed since they have run out of their gas and stopped star formation. This is done assuming an underlying stellar population synthesis (SPS) model and analyzing their spectral energy distributions (SED). Indeed, while redshift measurements can be obtained with quite high accuracy through the spectral line analysis, the age is not directly observable and one has to adopt particular techniques such as photometry, single spectral regions (e.g. D4000 \citep{Moresco:2020fbm}) or the full-spectral fitting.
Within these galaxies, old stellar populations have minimal star formation rates and this ensures low contamination from young components. However, such measurements are not immune to systematics and various criteria have been considered to minimize contamination from factors like photometry, spectroscopy, stellar velocity and mass dispersion. Potential degeneracies between physical parameters of the galaxy SED (e.g., age-metallicity) also contribute to the error budget. The reconstruction of the star formation history of age-dating galaxies is very challenging. Nevertheless, a strength of CCH lies in estimating the quantity $dt$ rather than the absolute age $t$. This approach minimizes the impact of systematic errors in absolute age estimation when measuring the differential age. All the possible sources of errors discussed above are taken into account in the covariance matrix provided by \cite{Moresco:2020fbm}, which includes the contribution of both statistical and systematic errors, $C^{\rm CCH}_{ij} = C^{\rm stat}_{ij} + C^{\rm sys}_{ij}$. 
In particular, in addition to the statistical error, the diagonal component of this matrix takes into account several systematics related to the estimate of the physical properties of a galaxy (e.g. stellar metallicity and the contamination by a younger stellar population), which are uncorrelated for objects at different redshifts. The non-null correlations arise from factors that instead rely on the common SPS model which, as explained above, is involved in studying the evolving galaxies.
Beyond General Relativity and the assumption that standard physics applies in the environment of the galaxy stars, CCH are devoid of additional cosmological assumptions. For this reason, in the last years, they emerged as suitable candidates for model-independent analyses. They can be employed in the calibration of BAO and SNIa \citep{Heavens:2014rja,Verde:2016ccp,Yu:2017iju,Gomez-Valent:2018hwc,Haridasu:2018gqm,Gomez-Valent:2021hda, Dinda:2022vmb,Favale:2023lnp,Mukherjee:2024akt} or other objects like GRBs or QSOs (see, e.g., \cite{Amati2019, Montiel, luongo2021model, luongo_intermediate_2022, Dinda:2022vmb,kumar_gamma_2023}).

In this work, we use state-of-the-art data on CCH, constituted of 33 data points in the redshift range $0.07<z<1.965$, to calibrate the GRB Dainotti correlations. The complete list of CCH and the corresponding references can be found in  \ref{app:data}. The covariance matrix of the data is computed as explained in \citep{Moresco:2020fbm}, using the codes provided in\footnote{\url{https://gitlab.com/mmoresco/CCcovariance}}.

\section{Methodology}\label{sec:method}

One can use measurements of the Hubble function to reconstruct the universe expansion history and solve the integral in Eq. \eqref{eq:DL} at the GRB redshifts. For this purpose, we employ GP. It is a machine learning tool that offers means of deriving cosmological functions through data-driven reconstructions under minimal model assumptions, keeping track of the correlations between them. Because of that, this algorithm has gained significant prominence in the context of model-agnostic regression techniques within the field of cosmology.

Being a generalization of a multivariate Gaussian, a Gaussian Process can be written by specifying its mean function $\bar{y}(z)$ and the covariance matrix $D(z, \tilde z)$, i.e, $f(z) \sim {\rm GP} (\bar{y}(z), D(z, \tilde z))$ \citep{2006gpml.book.....R}. If we denote $Z$ as the exact locations of the input data points, the covariance matrix $D$ takes the following form
\begin{equation}
D(z,\tilde z)\equiv\Biggl \{
    \begin{array}{lcl}
        K(z,\tilde z) + C(z,\tilde z) & {\rm if}\,z\,{\rm and}\,\tilde{z} \in Z\\
        K(z, \tilde z) & {\rm otherwise} \\
    \end{array}\,,
\end{equation}%
where $C$ is the covariance matrix of the data and $K(z,\tilde z)$ the kernel function. Indeed, although GP allow for agnostic (cosmology-independent) reconstructions, a specific kernel has to be chosen for the training. The latter is in charge of controlling the correlations between different points of the reconstructed function. Kernels depend on a set of hyperparameters, which follow a likelihood that is set by the probability of the GP to produce our data set at every point of the hyperparameter space. Before the reconstruction can take place, it is essential to determine the shape of this distribution. In many cases, this likelihood is sharply peaked and using the best-fit values of the hyperparameters becomes a good approximation \citep{Seikel:2012uu}. However, from a Bayesian perspective, the correct approach involves obtaining the complete distribution of the hyperparameters to account for their correlations and propagate their uncertainties to the final reconstructed function \citep{Gomez-Valent:2018hwc,Hwang:2022hla}. This is the approach followed in this work.

From every set of hyperparameters drawn from the hyperparameter distribution one can construct the GP at the locations $Z^\star \ne Z$, which is characterized by the mean function

\begin{equation}
    \bar f^{\star}=\bar{y}^{\star}+ K(Z^{\star},Z)[K(Z,Z)+C(Z,Z)]^{-1}(Y-\bar{y})\,
\end{equation}%
and covariance matrix
\begin{equation}\label{eq:GPcov}
   {\rm cov}(f^{\star})= K(Z^{\star},Z^{\star})-K(Z^{\star},Z)[K(Z,Z)+C]^{-1}K(Z,Z^{\star})\,,
\end{equation}%
where $Y$ are the data points located at $Z$ and $\bar{y}^{\star}\equiv\bar{y}(Z^\star)$ is the a priori assumed mean of the reconstructed function at $Z^\star$. From this GP, one can then produce a sufficiently large sample of the function of interest at $Z^\star$ to ensure convergence.

In this work, we reconstruct $H(z)$ at the redshifts $z_{\rm GRB}(=Z^{\star})$ of the $n_{\rm GRB}=20$ GRBs of the Platinum sample that fall below the highest CCH redshift, $z=1.965$,  employing an a priori zero mean function, i.e. $\bar{H}(Z^\star)=0$. The GRBs of our sub-sample lie therefore in the same redshift range of the calibrators. From the GP we obtain $N_{\rm real}=5\cdot10^{4}$ realizations of $H(z)$ at each $Z^{\star}$ and the same number of luminosity distances, by solving the integral in Eq. \eqref{eq:DL}. We ensure that the number of points in which we evaluate this integral is enough to have an accurate determination of the reconstructed function. This test is especially important in the region of redshifts below the GRB data point with the lowest redshift in our sample. To this purpose, we repeat the GP reconstruction of $H(z)$ by increasing the number of $Z^{*}$ in $0<z<0.553$ and find that the resulting shape of $\log D_L(z)$ remains stable. Thus, we are allowed to use the result obtained at $Z^{*} = z_{\rm GRB}$ to calibrate the Dainotti relations, Eqs. \eqref{eq:3DDainotti} and \eqref{eq:2DDainotti}. Indeed, combining Eq. \eqref{eq:3DDainotti} and \eqref{eq:lpeak} one can easily obtain the following quantity
\begin{equation}\label{eq:logDL3D}
\begin{split}
     \log D^{\rm{th}}_L= & a_{1}\log T^*_{\rm X} + b_{1}(\log F_{\rm peak} + \log K_{\rm prompt}) + c_1 +\\
     & +d_{1}(\log F_{\rm X} + \log K_{\rm plateau})\,,
\end{split}
\end{equation}%
which we denote as the theoretical value for the log-luminosity distance, with $D_L^{\rm th}$ written in units of cm. Here, $a_1 = a/2(1-b)$, $b_1 = b/2(1-b)$, $c_1 = ((b-1)\log(4\pi) + C_o)/(2(1-b))$ and $d_1 = -(1-b)/2$. 

On the other hand, the GP+CCH result can be used to compute the logarithm of the luminosity distances, which we treat as our observed value, $\log D^{\rm obs}_L(z)$. We have checked that its distribution is Gaussian in very good approximation. The corresponding covariance matrix can be obtained as follows,
\begin{equation}\label{eq:C_obs}
C_{\rm obs, ij}=\frac{1}{N_{\rm real}}\sum_{\mu=1}^{N_{\rm real}}(x_{\mu,i}-\bar{x}_i)(x_{\mu,j}-\bar{x}_j)\,,   
\end{equation}%
where $x_{\mu,i}$ is the value of $\log D^{\rm obs}_L(z)$ at the $i$-th redshift for each realization $\mu=1,...,N_{\rm real}$.

If we first consider uncorrelated errors, $\sigma_{\rm obs,i}$, i.e., if we consider a diagonal covariance matrix, we can build a chi-squared which takes the following form,

\begin{equation}\label{eq:chi2}
    \chi^{2}(a,b,C_o,\sigma_{int}) = \sum^N_{i=1} \frac{[ \log{D}^{\rm obs}_{L}(z_i) - \log{D}^{\rm{th}}_{L}(z_i,a,b,C_o)]^2}{(\sigma_{\rm obs,i}^{2} + \sigma_{int}^{2})}\,,
\end{equation}%
where $N = n_{\rm GRB}$. We sample the parameters of interest of the Dainotti relations $a, b, C_o$, and $\sigma_{int}$ through a Monte Carlo Markov Chain (MCMC) analysis, making use of the \texttt{Python} public package \texttt{emcee}\footnote{\url{https://emcee.readthedocs.io/en/stable/}} \citep{Foreman_Mackey_2013}, an implementation of the affine invariant MCMC ensemble sampler by \cite{Goodman2010}.
The log-likelihood to be evaluated at each step of the Monte Carlo is 
\begin{equation}\label{eq:like}
    \begin{split}
        \ln {\mathcal{L}} &=  -\frac{1}{2}\left [\sum^N_{i=1} \ln\{ 2\pi(\sigma^{2}_{\rm{obs},i}+\sigma^2_{int})\} +  \chi^{2} \right ] = \\
    &= -\frac{1}{2} \sum^N_{i=1} \ln \{2\pi(\sigma_{\rm obs,i}^{2} + \sigma_{int}^{2}) \}-\\ &-\frac{1}{2} \sum^N_{i=1} \frac{[ \log{D}^{\rm obs}_{L}(z_i) - \log{D}^{\rm{th}}_{L}(z_i,a,b,C_o)]^2}{\sigma_{\rm obs,i}^{2} + \sigma_{int}^{2}}\,.
    \end{split} 
\end{equation}%

In comparison to previous studies, we also allow the five parameters defining the fundamental plane ($\log T^*_{\rm X}, F_{\rm peak}, K_{\rm prompt}, \log F_{\rm X}, K_{\rm plateau}$, cf. Table \ref{tab:20platinum}) to vary freely in the Monte Carlo. We treat them as nuisance parameters. These parameters are specific for each GRB. This means that, in the end, if we employ the 3D Dainotti relation (Eq. \eqref{eq:3DDainotti}) we have $5\times n_{\rm GRB} = 100$ nuisance parameters. Of course, in the case of the 2D Dainotti relation (Eq. \eqref{eq:2DDainotti}), we only sample $\{\log T^*_{\rm X},\log F_{\rm X}$, $K_{\rm plateau}\}$, so we deal with $3\times n_{\rm GRB} = 60$ nuisance parameters. The posterior distribution is given by the product of the above likelihood and the Gaussian priors, and the constraints for the parameters of interest $a, b, C_o$, and $\sigma_{int}$ are derived after performing the marginalization over the nuisance parameters.

Due to the high dimensionality of the problem, to ensure convergence we choose to use a sufficiently large number of walkers ($\sim 220$ for the 3D relation and $\sim 140$ for the 2D one) as well as a large number of steps in the parameter space $n_{\rm steps} = 6\cdot 10^{5}$. We make use of the \texttt{Python} package \texttt{GetDist} \citep{Lewis:2019xzd} to obtain all the 1D posteriors and 2D contour plots shown in this paper, as well as the constraints for each parameter.

Given the novelty of our approach and in anticipation of future applications, in this work we evaluate the proposed methodology by assessing the impact of:

\begin{enumerate}
    \item  The prior choice in the Monte Carlo analyses. Given our aim to develop a method that is as model-independent as possible, we extensively test both Gaussian and flat priors to verify the robustness and compatibility of the results obtained with each.

First, we use Gaussian priors on the parameters of interest $a, b, C_o$, and $\sigma_{int}$. To build the priors we use the mean values and the 1$\sigma$ uncertainties from the fundamental plane fitting analysis for the X-ray Platinum sample with 20 GRBs (see Sec. \ref{sec:platinum}). Their values are listed in Table \ref{tab:prior_3D}. Then, we gradually increase the standard deviation of the Gaussian to 2-, 3- and 5$\sigma$, to reduce the informative imprint of the prior. Nonetheless, it is important to mention that to preserve the physical meaning of these parameters, we impose bounds on $a,b$ and $\sigma_{int}$. In particular, imposing $a<0$ ensures that there is an anti-correlation between $L_X$ and $T^*_X$ and when the value of $a$ is close to $-1$ this implies that the energy reservoir of the plateau remains constant (see Sec. \ref{sec:dainotticorr}). However, we will show that this anti-correlation is indeed maintained regardless of the physical cut $a<0$, as positive values of $a$ are consistently distant from the region preferred by the data in all the calibration results presented in this work. Regarding the $b$ parameter, we impose the prior $b>0$, which is driven by observational evidence and theoretical predictions. A negative $b$ is not physically supported, as it would mean that the higher the peak luminosity, the less bright the plateau. Indeed, as shown in, e.g., \cite{Dainotti11b,Hascoet:2014ira, dainotti2015b, Dainotti_2017}, there is observational evidence that the more kinetic energy there is in the prompt, the more is transferred in the afterglow phase. Thus, there is a positive correlation between $L_X$ and $L_{\rm peak}$, which supports the choice of the prior.
In addition, we will see that the use of this prior is also motivated by the strong degeneracy between $b$ and $C_0$, which cannot be broken with current data within this model-independent approach. Imposing $b>0$ allows us to break this degeneracy and thus avoid ending up in unphysical regions of the parameter space. Finally, the intrinsic scatter has to be greater than zero by definition. Lastly, we adopt flat priors on the parameters of interest, by transforming the previous Gaussian priors into uniform distributions. This conversion can be done in several ways and by applying different definitions. We decided to use the same approach employed by \cite{dainotti2023gamma}\footnote{Therein the authors test the impact of the prior choice in a similar way but to infer cosmological parameters with SNIa and BAO.}, which requires
    \begin{equation}\label{eq:transf_unif}
    \begin{aligned}
         &\mu = \frac{x_1+x_2}{2}\,, \\
         &\sigma = \sqrt{\frac{(x_2-x_1)^{2}}{12}}\,.
    \end{aligned}
    \end{equation}%
    
    These are the relations to transform a Gaussian distribution with mean $\mu$ and standard deviation $\sigma$ to a uniform distribution $\mathcal{U}(x_1,x_2)$ with the same mean and standard deviation.
    
    \item The covariance matrix of the reconstructed function $\log{D}^{\rm{obs}}_{L}(z)$ at the GRB redshifts, $C_{\rm obs}$, as defined in Eq. \eqref{eq:C_obs}. 
    The total covariance matrix, $C_{\rm T}$, incorporates also the contribution of the intrinsic scatter $\sigma_{\rm int}$, which is the same for all the GRB data points. Its elements read  $C_{\rm T,ij} = C_{\rm obs,ij} + \delta_{ij}\sigma_{int}^2$. Hence, if we define $\Delta_i \equiv \log{D}^{\rm{obs}}_{L}(z_i) - \log{D}^{\rm th}_{L}(z_i)$, generalizing Eq. \eqref{eq:like}, we can write the new log-likelihood
    
    \begin{equation}\label{eq:like_multi}
    \ln{\mathcal{L}} = -\frac{N}{2}\ln(2\pi) - \frac{1}{2} \ln\det(C_{\rm T}) - \frac{1}{2}\sum^{N}_{i,j=1} \Delta^{T}_i (C^{-1}_{\rm T})_{ij}\Delta_j\,,     
    \end{equation}%
    which duly takes into account the correlations.
\end{enumerate}%

\section{Results}\label{sec:results}
In this section, we present the results of the calibration of the GRB correlation parameters with CCH. Specifically, in Sec. \ref{sec:DL_GP}, we show how we can obtain the luminosity distance at the GRB redshifts without adopting any cosmological model, making use of the GP algorithm. Then, in Secs. \ref{sec:calib3D} and  \ref{sec:evo}, we use this result to perform the calibration of the 3D Dainotti correlation, without considering and considering the redshift evolution of the coefficients, respectively. These results lead us to study, in Sec. \ref{sec:calib2D}, the 2D Dainotti correlation, defined in terms of the X-ray time and luminosity at the end of the plateau emission. In all the aforementioned analyses, we investigate the impact of the prior choice on our results, drawing interesting conclusions about the calibrating role of the CCH, which we then discuss in detail in Sec. \ref{sec:discussion}. In Sec. \ref{sec:ext_ladder}, we finally present the extension of the distance ladder obtained from the calibrated GRBs.

\subsection{GRB luminosity distance from CCH and Gaussian Processes}\label{sec:DL_GP}

\begin{figure}[t!]
    \centering
    \includegraphics[width=\columnwidth]{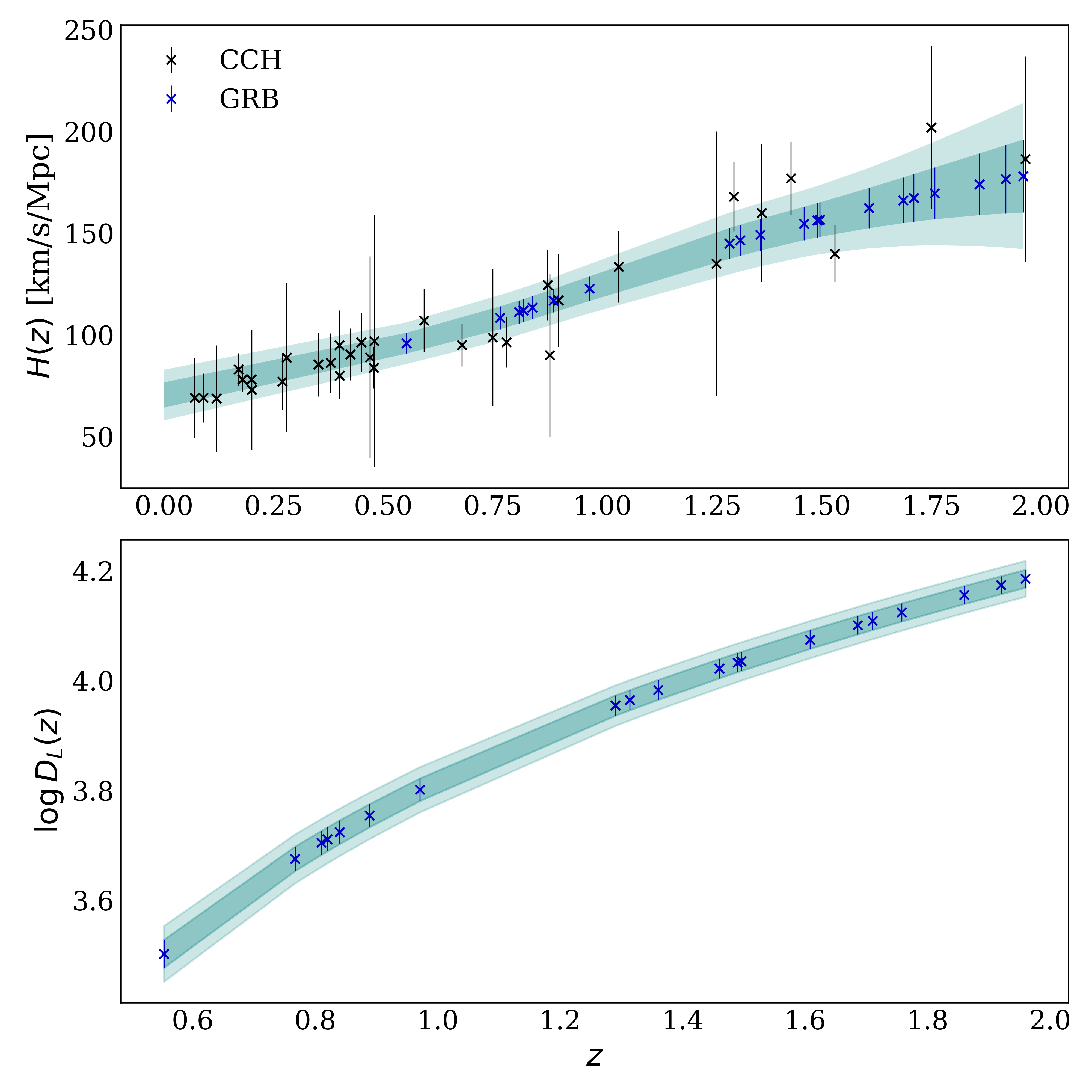}
    \caption{\textit{Top panel:} Reconstruction at 2$\sigma$ C.L. of $H(z)$ with GP. We show the CCH data (in black) and the individual values of $H(z)$ at the various GRB redshifts (in blue), both at 1$\sigma$ C.L. \textit{Bottom panel:} The corresponding shape of $\log{D_L(z)}$. The argument of the $\log$ is made dimensionless by dividing the luminosity distance by  1 Mpc.}\label{fig:HZDL}
\end{figure}

 We first reconstruct the Hubble function using the CCH data and the associated covariance matrix described in Sec. \ref{sec:cch}, making use of the public package {\it Gaussian Processes in Python} (\texttt{GaPP}) \footnote{\url{https://github.com/carlosandrepaes/GaPP}}, first developed by \cite{Seikel:2012uu}. We adopt the Matérn32 kernel and obtain the full distribution of its hyperparameters with \texttt{emcee}, as explained in Sec. \ref{sec:method}. This is something that has not been done in previous works that employ the GP technique to calibrate GRB correlations (either with SNIa or CCH), and this is certainly important to properly compute the uncertainty of the reconstructed function. This also applies to the use of the CCH covariance matrix. Neglecting the CCH correlations in the GP reconstruction has a modest impact on the final shape of $H(z)$, as we have explicitly checked. The relative difference in units of $\sigma$ between the correlated and uncorrelated CCH reconstructions is at most 15\%, dropping at 1\% in the data-poor region, for $z\gtrsim1$ (with $\sigma$ being the largest uncertainties of the two reconstructions, i.e. the CCH-correlated one). Regarding the selection of the kernel, in \cite{Favale:2023lnp}, it is shown that different (stationary) kernels provide very similar results when the same CCH data set is used to reconstruct $H(z)$ and also that the most conservative choice is the Matérn32 covariance function, since it is the one that leads to the largest error budget, see the aforesaid reference for details. Supported by these previous results, we follow the same approach here.
We reconstruct the Hubble function within the CCH data redshift range, i.e. $z\leq 1.96$. The resulting shape of $H(z)$ is shown in Fig. \ref{fig:HZDL}, together with the corresponding result for $D_L(z)$, obtained by using the GP-shapes of $H(z)$ and Eq. \eqref{eq:DL}. 

\subsection{Calibration of the 3D Dainotti relation}\label{sec:calib3D}

\begin{figure}
    \centering
    \includegraphics[width=\columnwidth]{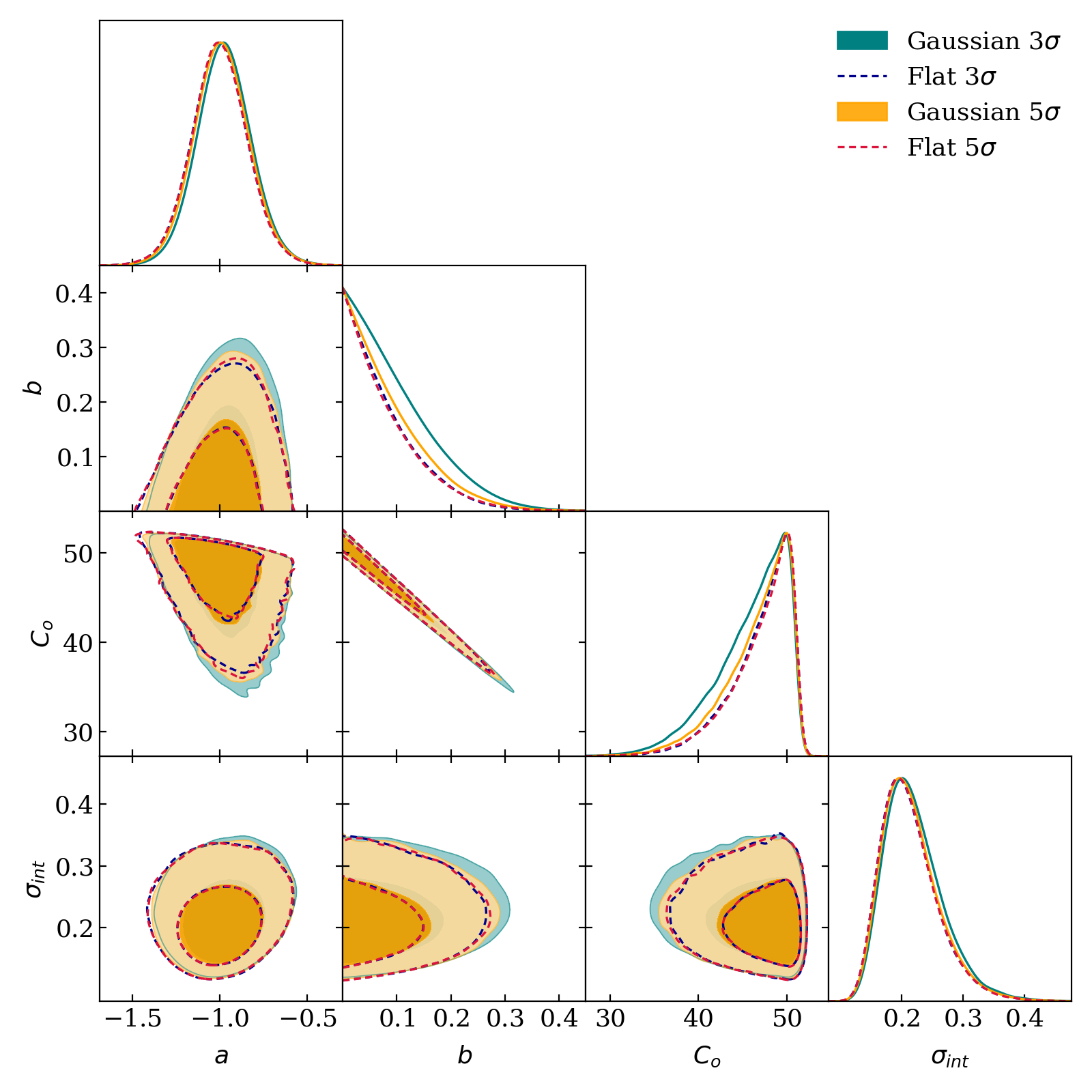}
    \caption{1D posteriors and 2D contour plots at 68\% and 95\% C.L. obtained in the analysis of the 3D Dainotti relation using Gaussian priors and the corresponding flat priors from Eq. \eqref{eq:transf_unif}  at 3- and 5$\sigma$. We use dashed lines for the results obtained with flat priors. They match almost perfectly with those obtained with Gaussian priors.}

    \label{fig:triangle_3D}
\end{figure}

Once obtained the reconstruction of $\log D^{\rm obs}_L(z)$ together with its uncertainties, we can constrain the correlation parameters in Eq. \eqref{eq:3DDainotti} as well as the intrinsic scatter of the plane, by means of the likelihood in Eq. \eqref{eq:like}.
As already mentioned in Sec. \ref{sec:method}, to ensure the efficiency of the method and cross-check the stability of the results, we need to quantify the impact of the prior choice on the parameters of interest $a, b, C_o$, and $\sigma_{int}$ in the Monte Carlo sampling.
We thus employ Gaussian priors with a mean equal to that obtained in the fundamental plane (FP) fitting analysis and a standard deviation equal to 1, 2, 3 or 5 times the one obtained in the FP analysis (cf. Table \ref{tab:prior_3D}). We present our results in Fig. \ref{fig:3D_gauss}. We then repeat the analysis making use of flat priors. We apply the conversion from Gaussian to flat priors of Eq. \eqref{eq:transf_unif}. However, in Table \ref{tab:prior_3D}, we only report the results of this conversion for the 3- and 5$\sigma$ cases since they lead to the most conservative results. We show the corresponding MCMC outcome in Fig. \ref{fig:triangle_3D}. Some conclusions follow:

\begin{figure}[t!]
    \centering
    \includegraphics[width=\columnwidth]{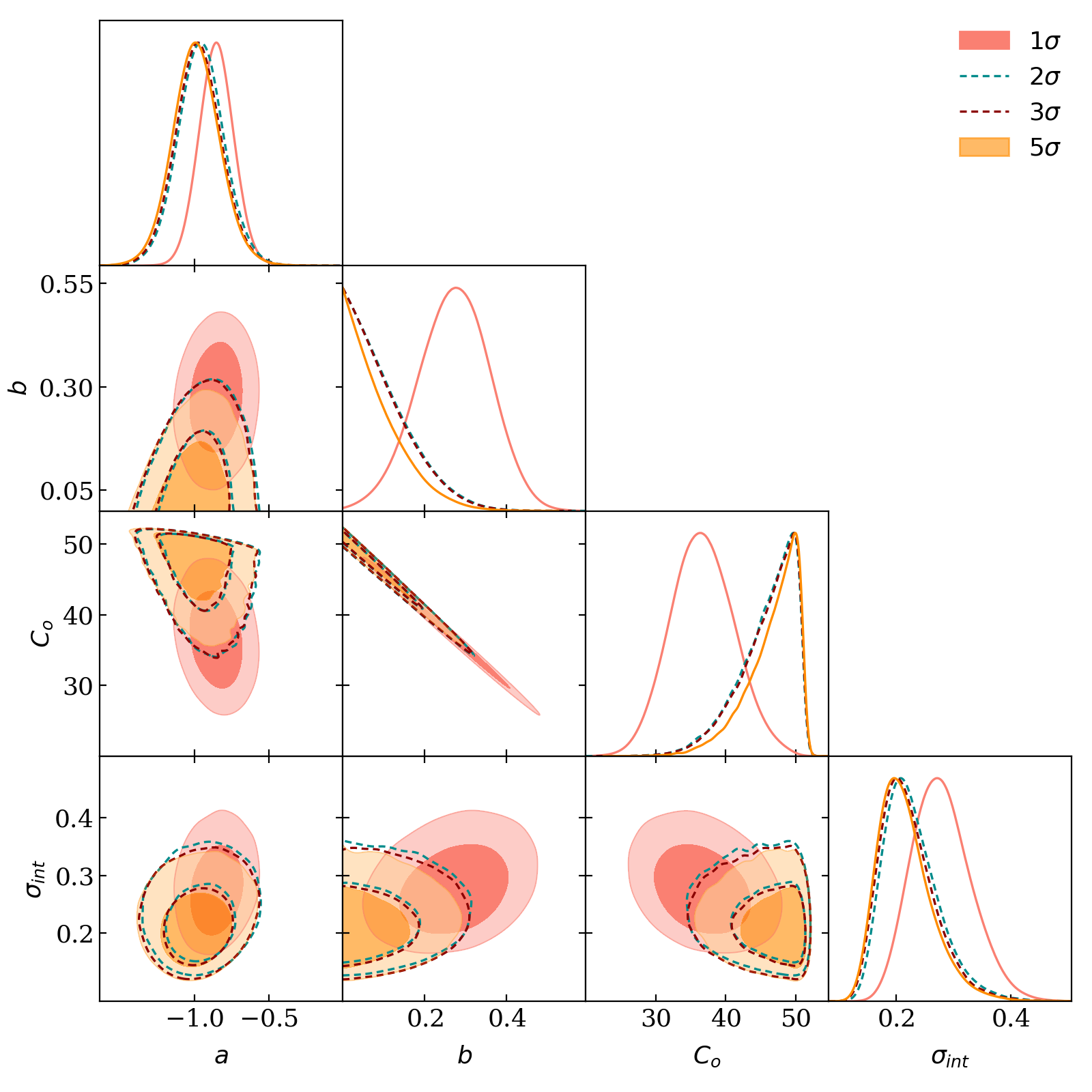}
    \caption{1D posteriors and 2D contours at 68\% and 95\% C.L. obtained in the analyses of the 3D Dainotti relation, assuming the \textit{n}-$\sigma$ progression for Gaussian priors and the physical cuts $a<0$, $b>0$, $\sigma_{int}>0$. Results obtained in the different cases are all consistent with each other. However, the 2$\sigma$, 3$\sigma$ and 5$\sigma$ priors are more conservative and allow for more model-independent results.}  
    \label{fig:3D_gauss}
\end{figure}%

\begin{table}[t!]
  
\centering
\begin{tabular}{l c c c}          
\hline\noalign{\smallskip}        
& $ \rm FP_{\rm BF}$ & $ 3\sigma$  & $ 5\sigma$  \\   
\hline\noalign{\smallskip}                                 
   a & $-0.81\pm0.17$ & (-1.68, 0) & (-2.26, 0)\\
   b &$ 0.50\pm0.17$ & (0, 1.37) & (0, 1.95)\\
   $C_o$ &$ 24.65\pm8.91$ & (-22,71) &(-53,102)\\
   $\sigma_{int}$ &$ 0.35\pm0.07 $&   (0, 0.72)& (0, 0.97) \\
\hline                                            
\end{tabular}
\caption{Priors adopted for the 3D relation analysis. In the first column, we report the mean values and the corresponding 1$\sigma$ uncertainties obtained from the GRB fundamental plane fitting (FP$_{\rm BF}$) analysis, from which we take the Gaussian priors at \textit{n-}$\sigma$. In the second and third columns, we list, respectively, the uniform prior ranges at 3- and 5$\sigma$ C.L., computed using Eq. \eqref{eq:transf_unif}. We impose physical cuts on the parameters ($a<0$, $b>0$, $\sigma_{int}>0$). This is why some of the boundaries in the last two columns coincide.}
  \label{tab:prior_3D}
\end{table}%

\begin{itemize}
    \item Fig. \ref{fig:3D_gauss} shows that results obtained with different Gaussian priors are all compatible with each other within errors. The largest differences are found with the 1$\sigma$ prior. 
    This difference can be easily understood. In the 1$\sigma$ prior case, we only let the walkers explore a very limited region of the parameter space around the fundamental plane mean values.
    
    \item Interestingly, as soon as we move away from the 1$\sigma$ prior and the priors become less informative, the posteriors for $a, b, C_o$, and $\sigma_{int}$ do not significantly change between different prior choices. This suggests that the method is stable. This is even more evident if we look at Fig. \ref{fig:triangle_3D}, which shows that passing from Gaussian to uniform priors does not alter the posteriors significantly.
    Given the stability of the results, hereafter, we will consider the set-up with flat priors at 5$\sigma$ as our \textit{baseline}, being this the most conservative choice.
    We report the mode and mean values of the one-dimensional marginalized distributions for each of the parameters of interest in Table \ref{tab:results_5s}.
    
    \item We find no preference for a non-zero value of $b$, suggesting that this parameter is not playing a major role in the evaluation of the cosmological results and hence we can use the 2D relation instead. Actually, its upper bound at 95\% C.L. stands around $0.2$ in all the analyses presented so far. From a Bayesian perspective, we are therefore motivated to consider the 2D Dainotti relation, given by Eq. \eqref{eq:2DDainotti}. We will study this scenario in Sec. \ref{sec:calib2D}.
    
    \item There is a strong degeneracy between $C_o$ and $b$. The one-dimensional posterior of the normalization factor exhibits a cut-off around $C_o\sim 52$, which is due to the aforementioned correlations and the physical bound $b>0$ employed in our analysis of the 3D Dainotti relation (see Sec. \ref{sec:method}). Indeed, in the 2D case we consistently find $C_o\sim 51$, since we are essentially saturating the prior by setting $b=0$, cf. Sec. \ref{sec:calib2D}.
 
\end{itemize}%

\begin{table*}[t!]
\centering
\begin{tabular}{p{0.15\linewidth}p{0.15\linewidth} c c c c}          \hline\noalign{\smallskip}        
 & & a & b & $C_o$  & $\sigma_{\rm int}$ \\
\hline\noalign{\smallskip}
     &Mode values & $-1.03$  & $<0.21 $& $50.17$  & $0.20$\\
     3D relation &&&& \\
    & Mean & $-1.00\pm0.16$  & $<0.21$ & $47.05_{-1.35}^{+4.21}$ & $0.21_{-0.05}^{+0.03}$ \\
      \noalign{\smallskip}
    \hline
    \noalign{\smallskip}
    &Mode values & $-1.07$  & & $51.19$ & $0.19$\\
     2D relation &&&& \\
   & Mean & $-1.04\pm0.16$  & & $51.16\pm0.53$ & $0.21_{-0.05}^{+0.03}$ \\
   \noalign{\smallskip}
\hline                                            
\end{tabular}
\caption{Constraints on the parameters of interest for the 3D and 2D Dainotti relations sampled assuming the baseline set-up, i.e. uniform priors at 5$\sigma$ C.L., which leads to the most conservative results in this paper (see also Figs. \ref{fig:triangle_3D} and \ref{fig:2D}). We report for each parameter the mode and mean value together with its associated uncertainty at 68\% C.L., except for $b$, for which we give an upper bound at 95\% C.L.}\label{tab:results_5s}
\end{table*}%

\begin{figure}[t!]
    \centering
    \includegraphics[width=\columnwidth]{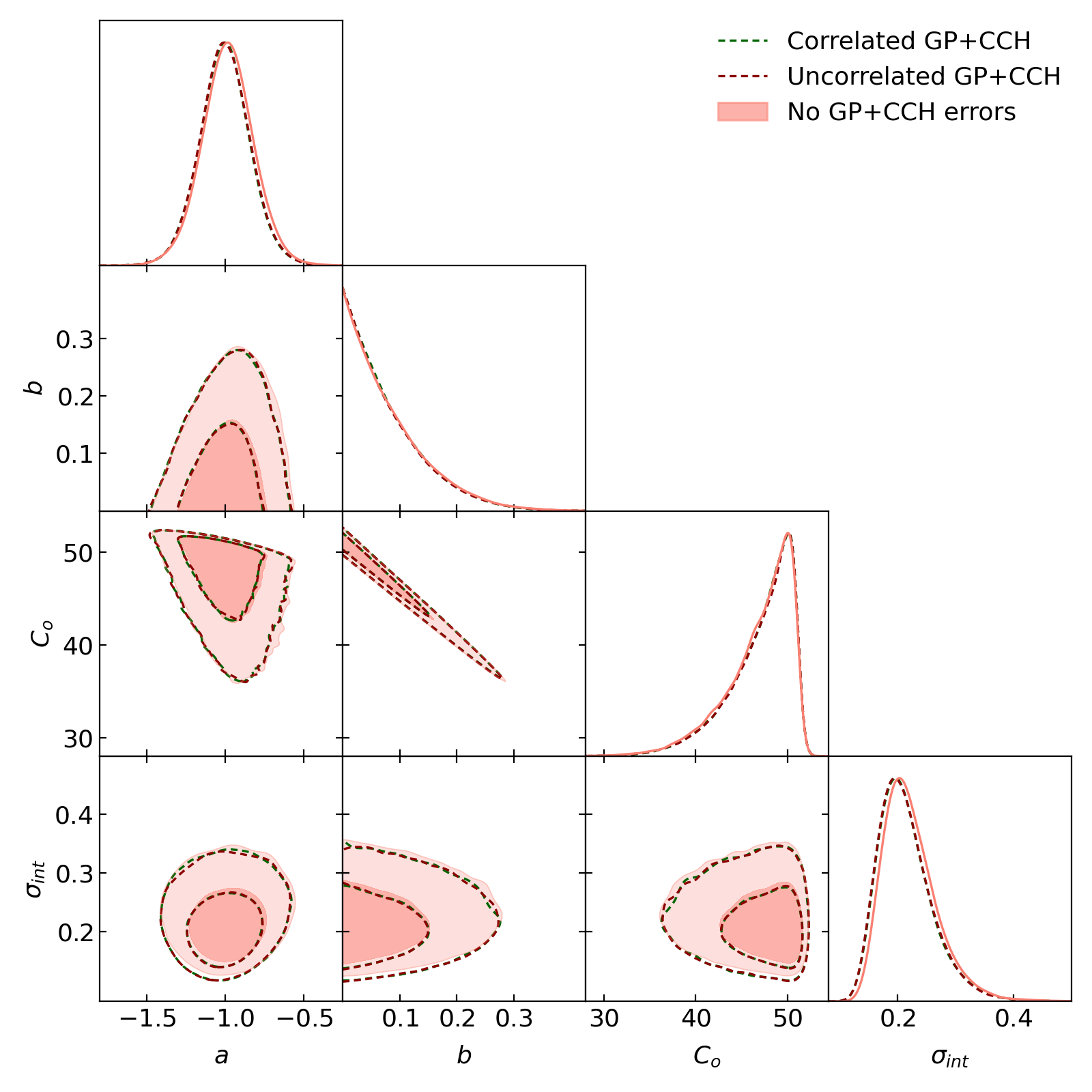}
    \caption{1D posteriors and 2D contour plots at 68\% and 95\% C.L. for the parameters of the 3D Dainotti relation, obtained assuming flat priors at 5$\sigma$. We study the impact of accounting for correlations between redshifts in the luminosity distance estimates reconstructed with GP+CCH (dotted green), neglecting those correlations (dotted red) and completely ignoring the uncertainties in the reconstruction (in pink). Differences are derisory.}
    \label{fig:3D_cov}
\end{figure}%

Finally, in order to quantify the impact of correlations between different redshifts in $\log D^{\rm obs}_L(z)$ obtained from GP+CCH, we perform an additional analysis in which we also include the contribution of the non-diagonal terms of the covariance matrix $C_{\rm obs}$ (Eq. \eqref{eq:C_obs}) by making use of the log-likelihood in Eq. \eqref{eq:like_multi}.
We compare the results obtained in this new analysis with those discussed above, obtained when we only consider a diagonal covariance matrix. Moreover, we also investigate the scenario in which we neglect the uncertainties of the reconstructed luminosity distances and only consider the effect of the intrinsic scatter, $C_{\rm T,ij} = \delta_{ij}\sigma_{int}^2$. The results are presented in Fig. \ref{fig:3D_cov}, illustrating that the posteriors for the parameters $a, b, C_o$, and $\sigma_{int}$ are insensitive to the GP+CCH uncertainties. In Sec. \ref{sec:discussion}, we will provide two interpretations of this conclusion.

\subsection{Calibration of the 3D Dainotti relation including redshift evolution corrections}\label{sec:evo}

To ensure the generality of our treatment, it is worth discussing the role of the corrections due to evolutionary effects. Indeed, each physical feature of the three-dimensional GRB fundamental plane, $L_{\rm X}$, $T^{*}_{\rm X}$ and $L_{\rm peak}$, is affected by selection biases due to instrumental thresholds and redshift evolution of the variables involved in the correlations. It is shown in \citep{Dainotti2013,Dainotti2015,dainotti2015b,dainotti2020x,Dainotti22, dainotti2022gamma} that in order to correct for these effects, one can employ the Efron \& Petrosian method \citep{efron1992simple}, which tests the statistical dependence among $L_{\rm X}$, $T^{*}_{\rm X}$ and $L_{\rm peak}$. For details, see also \cite{dainotti17a,dainotti2023gamma}. Once one introduces this correction, the expression for the GRB fundamental plane takes the following form:

\begin{equation}\label{eq:3D_evo}
\begin{split}
\log L_{\rm X} - k_{L_x}\log(1+z) = & a_{ev}(\log T^{*}_{\rm X} - k_{T^*_{\rm X}}\log(1+z))+\\ &+ b_{ev}(\log L_{\rm peak} - k_{L_{\rm peak}}\log(1+z)) +\\&+ C_{o,ev}\,.
\end{split}
\end{equation}%
The subscript \textit{ev} is employed here to distinguish the relation parameters from those employed in the non-evolutionary case, while $k_{L_x}, k_{T^*_{\rm X}}$ and $k_{L_{\rm peak}}$ represent the evolutionary coefficients related to each physical feature.

Starting from Eq. \eqref{eq:3D_evo}, one can derive the new likelihood, which now accounts for redshift evolution effects, thus assessing how they can affect the correlation parameters. We evaluate the impact of flat priors on $a_{ev}, b_{ev}, C_{o,ev}, \sigma_{int,ev}$ but only at 3- and 5$\sigma$, motivated by our previous results. To use the transformation in Eq. \eqref{eq:transf_unif}, we started from the results obtained from the analysis of the Platinum sample presented in \citep{dainotti2023gamma}, which, at 1$\sigma$, read: $a_{ev} = -0.85\pm0.12$, $b_{ev} = 0.49\pm 0.13$, $C_{o,ev}= 25.4\pm 6.9$, and $\sigma_{\rm int,ev} = 0.18\pm0.09$. For the evolutionary coefficients entering Eq. \eqref{eq:3D_evo}, we take advantage of two results obtained for the full Platinum sample (50 GRBs) and also for the Whole sample (222 GRBs) in \citep{dainotti2023gamma}. At 68\% C.L. they read, respectively:
\begin{equation}\label{eq:evo_coeff}
    \begin{aligned}
        &k_{L_x} = 1.37^{+0.83}_{-0.93} , \ \ k_{T^*_{\rm X}} = -0.68^{+0.54}_{-0.82} , \ \ k_{L_{\rm peak}} = 0.44^{+1.37}_{-1.76}\,, \\
        &k_{L_x} = 2.42^{+0.41}_{-0.74} , \ \ k_{T^*_{\rm X}} = -1.25^{+0.28}_{-0.27} , \ \ k_{L_{\rm peak}} = 2.24\pm0.30\,.
    \end{aligned}
\end{equation}%
\begin{figure}[t!]
\centering
\includegraphics[width=\columnwidth]{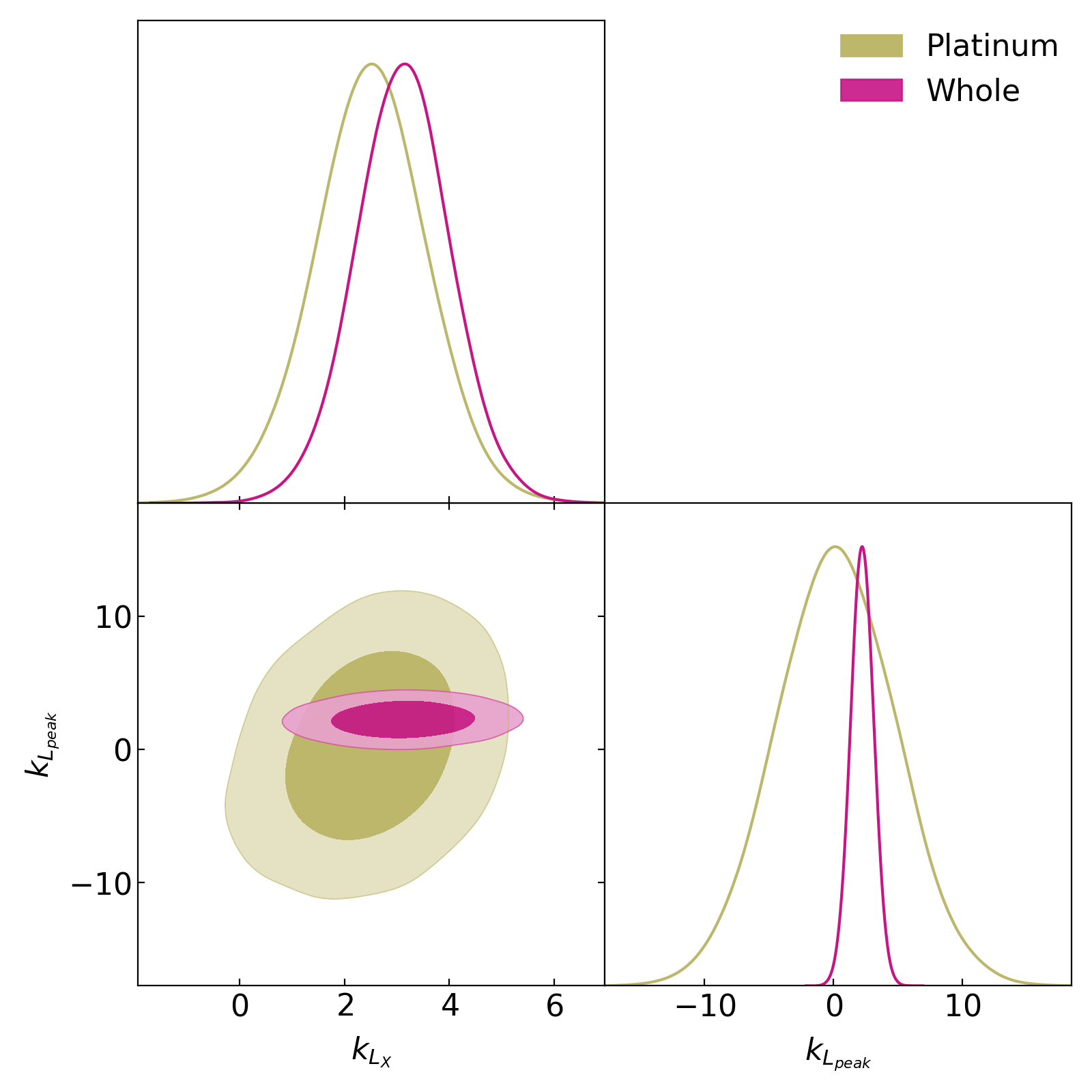}
    \caption{1D posteriors and 2D contour plots at 68\% and 95\% C.L. for the evolutionary coefficients $k_{L_{\rm X}}$ and $k_{L_{\rm peak}}$, sampled assuming Gaussian priors at 3$\sigma$ C.L. from the Platinum sample (in green) or from the Whole sample (in magenta). For the parameters $a_{ev}, b_{ev}, C_{o,ev}$ (not shown here) we impose flat priors at 5$\sigma$, cf. the main text and Fig. \ref{fig:whisker_evo} for details.}
    \label{fig:triangle_ev}
\end{figure}%

\begin{figure*}[t!]
    \centering
    \includegraphics[scale=0.6]{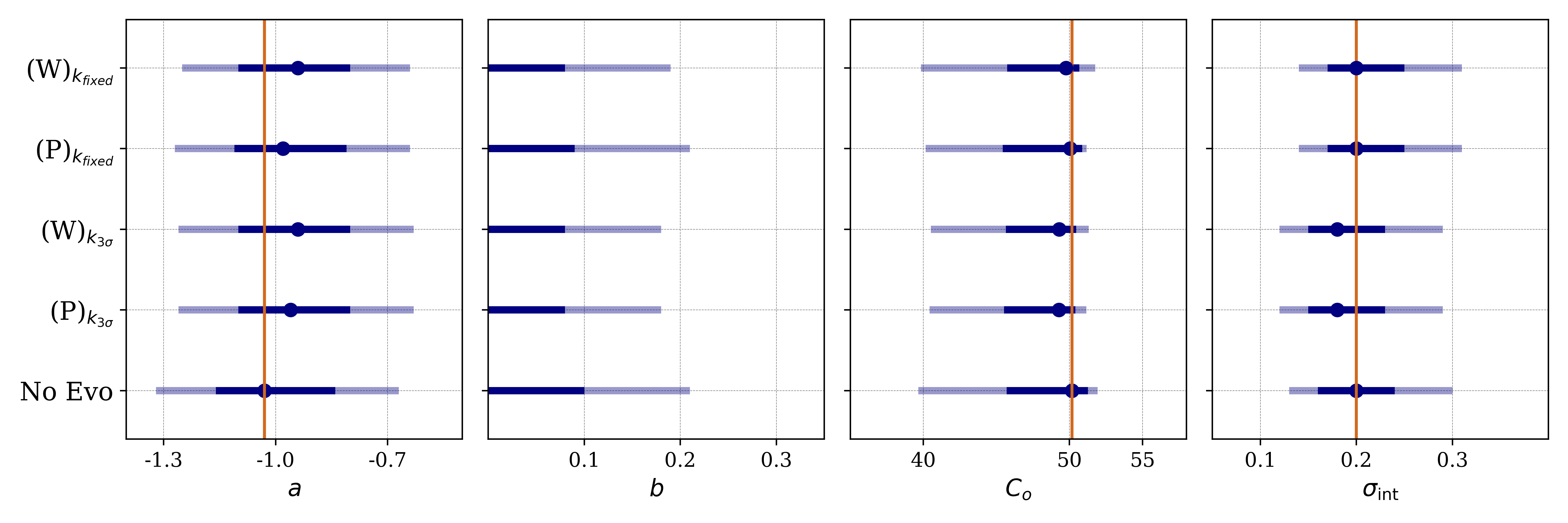}
    \caption{Whisker plot for the parameters of interest evaluated at 68\% and 95\% C.L. (for $b$ we only show its upper bounds), with the dots indicating the mode values of the one-dimensional marginalized distributions. The orange line passes through the \textit{No Evo} (no evolution) mode values, obtained from the analysis with flat priors at 5$\sigma$ in Sec. \ref{sec:calib3D}. With \textit{W} we denote the results obtained considering the evolutionary coefficients from the Whole sample, the Platinum is instead indicated with \textit{P}. The label $k_{fixed}$ refers to the analyses in which the $k$'s are fixed, whereas $k_{3\sigma}$ refers to those analyses in which the $k$'s are allowed to vary within $3\sigma$. For details, we refer the reader to the main text in Sec. \ref{sec:evo}. In the cases with evolution, $a, b, C_o, \sigma_{int}$ become, respectively, $a_{ev}, b_{ev}, C_{o, ev}, \sigma_{int, ev}$.}
    \label{fig:whisker_evo}
\end{figure*}
The above constraints are obtained assuming a flat $\Lambda$CDM model. We decide to proceed in two ways:
\begin{itemize}
\item Keep them fixed (Case 1);
\item Let them vary in the Monte Carlo analysis, treating them as nuisance parameters (Case 2).
\end{itemize}%
In Case 1), we just fix the $k$'s to the central values reported above. In Case 2), we also use their uncertainties to build the corresponding priors. This is equivalent to propagating them to the parameters of interest in the MCMC analysis. This additional case is indeed performed to coherently reduce the model dependency level with this work's investigation line. Notice, though, that since $k_{T^*_X}$ does not depend on cosmology (being related to a measure of a characteristic time scale for the end of the plateau emission), we keep it fixed while we treat as nuisance parameters $k_{L_{\rm X}}$ and $k_{L_{\rm peak}}$. Since, as one can appreciate from Eq. \eqref{eq:evo_coeff}, the Platinum constraints already come with large errors due to the small sample size and a large number of parameters, we decide to set Gaussian priors (instead of uniform priors) at 3$\sigma$ on both $k_{L_{\rm X}}$ and $k_{L_{\rm peak}}$ to avoid non-physical results for these parameters. We also tested what happens if we consider uniform priors at 3$\sigma$. We have checked that the results for $a_{ev}, b_{ev}$, and $C_{o,ev}$ are not affected by this choice.

In Fig. \ref{fig:triangle_ev}, we show the corresponding posteriors of the evolutionary coefficients. At 68\% C.L., we obtain $k_{L_{\rm X}} = 2.50_{-1.05}^{+1.06}$ and $k_{L_{\rm peak}}=0.20_{-4.66}^{+4.67}$ when we employ priors from the Platinum sample, while we find $k_{L_{\rm X}} = 3.12 \pm 0.90$ and $k_{L_{\rm peak}} = 2.23 \pm 0.89$ from the Whole sample. As expected, the uncertainties for the latter case are smaller since the fundamental plane fitting analysis for the Whole sample has more constraining power (see Eq. \eqref{eq:evo_coeff}), being the data set larger.
We here notice that from our analysis, similarly to the analysis performed by \cite{dainotti2023gamma} the parameters of the evolution are compatible within 1$\sigma$ for both $k_{L_{\rm X}}$, $k_{L_{\rm peak}}$ showing the reliability of this analysis as well.
For what concerns the parameters of interest, we prefer to show their constraints in a whisker plot in Fig. \ref{fig:whisker_evo}, to compare the results throughout all the different cases described above (i.e., Cases 1 and 2, and considering no evolution), using flat priors at 5$\sigma$ C.L.\footnote{We have also tested the case with uniform priors at 3$\sigma$ on $a_{ev}, b_{ev}, C_{o,ev}, \sigma_{int,ev}$. However, for a specific set of $k$'s (Whole or Platinum), the posteriors from 3- or 5$\sigma$ do not change. Therefore, we prefer to show only the comparison between the analyses for a specific prior range choice (in this case, 5$\sigma$).}.  It is remarkable to notice, again, that different settings provide very stable results on the correlation parameters, even when redshift evolution corrections are accounted for. Additionally, despite very small differences, all the scenarios lead to a tight relation, especially when the $k$'s vary freely. Indeed, in this case we find $\sigma_{int}=0.20^{+0.03}_{-0.05}$ instead of $\sigma_{int}=0.22^{+0.03}_{-0.05}$ (with mode values 0.18 instead of 0.20, respectively).
If we consider that $\sigma_{int}$ is a measure of the scatter, which can depend on the spin period and magnetic field variability in each GRB, values so close to each other mean that the system has similar values of these parameters, with small variations.

\subsection{Calibration of the 2D Dainotti relation}\label{sec:calib2D}

\begin{figure}[t!]
    \centering
    \includegraphics[width=\columnwidth]{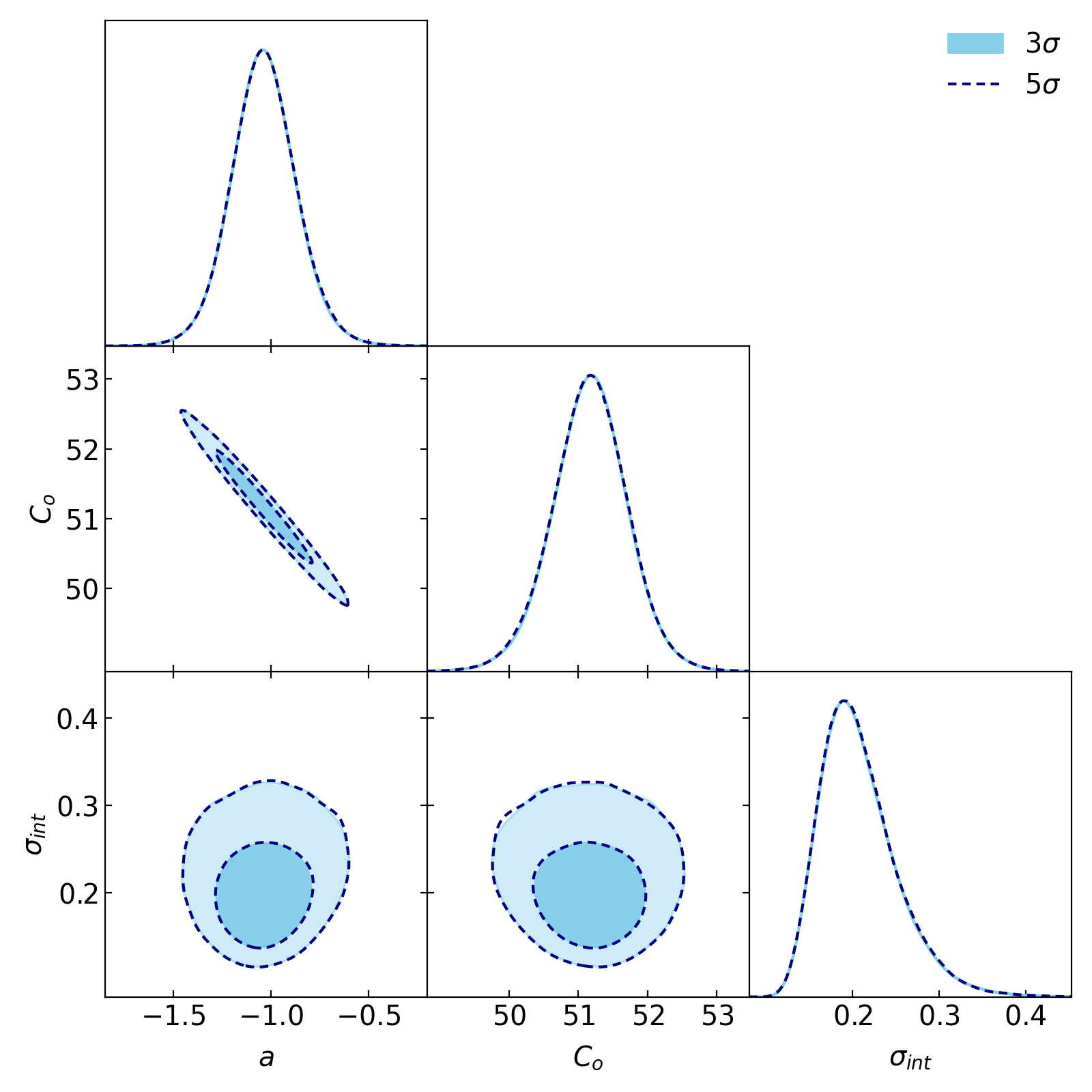}
    \caption{1D posteriors and 2D contour plots at 68\% and 95\% C.L. from the MCMC carried out for the 2D Dainotti relation assuming flat priors at 3- and 5$\sigma$, in cyan and blue, respectively. The results match each other, showing also in the case of the 2D relation the insensitivity of the analysis to the prior choice.}
    \label{fig:2D}
\end{figure}

As seen in Secs. \ref{sec:calib3D} and \ref{sec:evo}, the CCH calibration leads to a result for $b$ that is compatible with 0 at 68\% C.L. Thus, it is worth analyzing the 2D relation, which incorporates the properties of the end time of the plateau emission and the  X-ray luminosity. The parameters to sample are, in this case, $a, C_o$, and $\sigma_{\rm int}$.
Starting from Eq. \eqref{eq:2DDainotti}, the same reasoning used to obtain the log-luminosity distance in Eq. \eqref{eq:logDL3D} applies here, where we drop the contribution of $\log L_{\rm peak}$. It is straightforward to see that Eq. \eqref{eq:logDL3D} reduces to
\begin{equation}\label{eq:logDL2D}
        \log D^{\rm th}_L= a_{1}\log T^*_{\rm X} + c_1 + d_{1}(\log F_{\rm X} + \log K_{\rm plateau})\,,
\end{equation}%
where now $a_1 = a/2$, $c_1 = (-\log(4\pi) + C_o)/2$ and $d_1 = -1/2$. On the same line of Sec. \ref{sec:calib3D}, we assess the impact of the different priors in the MCMC analysis. Their values are listed in Table \ref{tab:prior_2D} while the MCMC results are shown in Fig. \ref{fig:2D}.

We report both the mode values and the means of the one-dimensional marginalized distributions in Table \ref{tab:results_5s}. The constraint we obtain for $a$ resonates well with the expectations: a slope of $\approx -1$ implies a constant energy reservoir during the plateau phase \citep{Dainotti2013, stratta2018magnetar}. In particular, we can ascribe this phase to energy injection by newly born neutron stars \citep{Dai:1998hm, Zhang:2000wx},  and this result supports that the rotational energy of the latter is constant, indicating that they can be treated as a standard candle \citep{Wang:2021hcx}. Here, however, we clarify that the construction of the Platinum sample starts from phenomenological choices of the flatness of the plateau, the absence of flares, etc. and therefore, in our approach, we can independently obtain the magnetar-driven parameters. Indeed, the $a$ slope is independent of the initial priors.

\begin{table}[t!]
\centering                                     
\begin{tabular}{l c c c}          
\hline\noalign{\smallskip}  
& $ \rm FP_{\rm BF}$ & $ 3\sigma$  & $ 5\sigma$  \\   
\hline\noalign{\smallskip}                             
   a & $-1.03\pm0.16$ &(-1.85, -0.2) &(-2.39, 0)\\      
   $C_o$ &$ 51.2\pm0.52$ & (48, 54) & (47,56)\\
   $\sigma_{int}$ &$ 0.43\pm0.08 $&  (0.01, 0.85)& (0, 1.13) \\
\hline                                            
\end{tabular}
\caption{Priors adopted for the 2D relation analysis. The first column reports the mean values and the corresponding 1$\sigma$ uncertainties obtained from the GRB fundamental plane fitting analysis. In the next two columns, we report uniform prior ranges at 3- and 5$\sigma$ C.L., respectively, computed from Eq. \eqref{eq:transf_unif}.}
\label{tab:prior_2D}
\end{table}%

\begin{figure}[t!]
    \centering
    \includegraphics[width=\columnwidth]{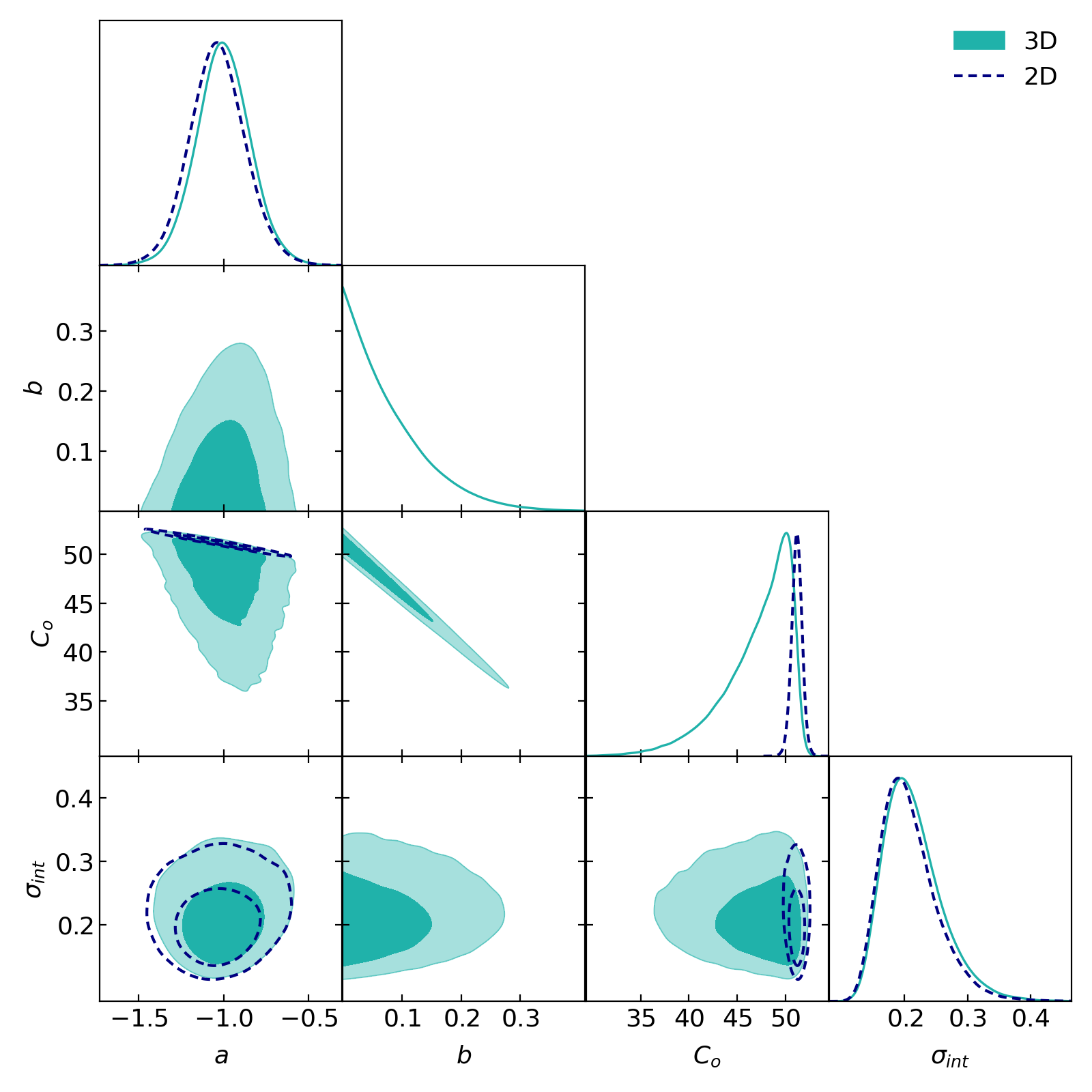}
    \caption{Comparison between the MCMC results for the 3D and 2D Dainotti relations obtained with the baseline set-up in Secs. \ref{sec:calib3D} and \ref{sec:calib2D}, respectively.}
    \label{fig:2D3D}
\end{figure}%

In the 2D Dainotti relation, we essentially set $b=0$, so we break the strong degeneracy between this parameter and $C_o$ (see Sec. \ref{sec:calib3D} and Fig. \ref{fig:2D3D}).
This explains why the uncertainty on $C_o$ decreases by 80\% compared to the baseline constraint obtained in Sec. \ref{sec:calib3D}. This fact may enhance our ability to constrain cosmology.

Finally, we assess the impact of the correlations in the values of $\log\, D_L^{\rm obs}$ obtained from the GP reconstruction by applying to the 2D relation the procedure explained at the end of Sec. \ref{sec:calib3D}. We find that our conclusions remain valid within this framework as well.

\subsection{The extension of the distance ladder with and without evolution}\label{sec:ext_ladder}

\begin{figure}[t!]
    \centering
    \includegraphics[width=\columnwidth]{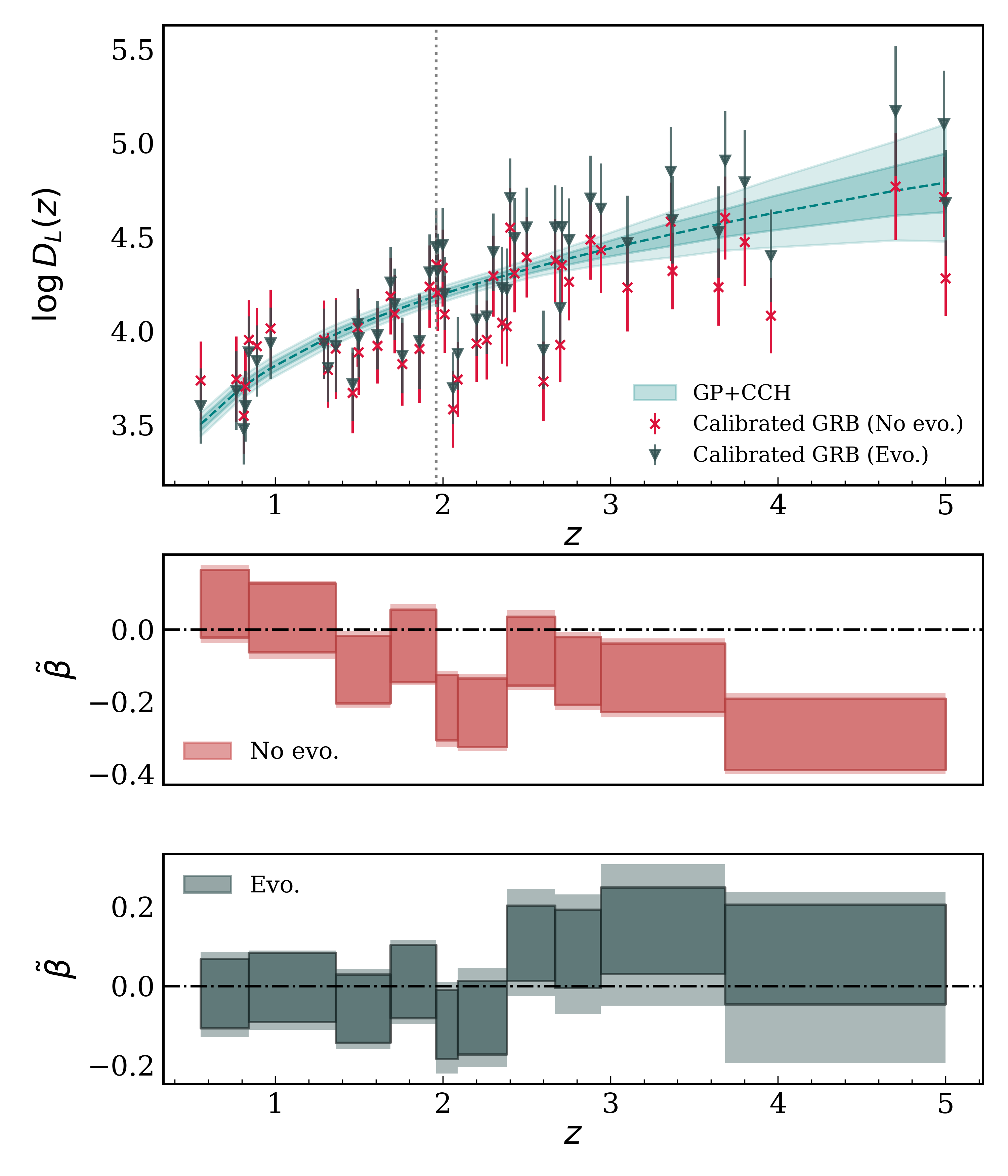}
    \caption{\textit{Upper plot}: The extended distance ladder. The GP+CCH reconstruction of the logarithm of the luminosity distance as a function of the redshift is shown in light green. The calibrated GRB data obtained with and without evolution effects are presented in grey and light red, respectively. The dotted vertical line at $z=2$ is used to mark the border between the data employed in the calibration process and those used only in the extension of the ladder. The error bars of the calibrated GRB distances stand at 1$\sigma$ C.L., while for the GP result we show both the $1\sigma$ and 2$\sigma$ bands. The luminosity distance is given in Mpc; \textit{Middle plot}: Weighted mean of the bias between the GP+CCH reconstruction and the calibrated GRB data with its uncertainty, as given in Eq. \eqref{eq:bias}, when the evolution effects are neglected. Lighter-coloured boxes are obtained when correlations are accounted for, see Sec. \ref{sec:ext_ladder} for details; {\it Bottom plot}: The same as in the middle plot, but considering the evolution effects.}
    \label{fig:ext_DL}
\end{figure}%

Supported by the stability of the results achieved in the previous sections, it becomes natural to ask how such results can be leveraged in cosmological applications. As already mentioned, the advantage of using probes such as GRBs lies in their ability to extend the ladder beyond SNIa and BAO redshifts. This is the reason why we now employ the calibration results obtained making use of the GRB data at $z\lesssim2$ to obtain luminosity distances from the remaining GRBs contained in the Platinum sample and cover the region up to $z=5$ (we refer the reader to Table A1 of \cite{cao_gamma-ray_2022} for these additional GRBs data). We first check that the correlation coefficients between the calibration and nuisance parameters of the 20 GRBs involved in the calibration process are negligible and also that the posterior distributions of the nuisance parameters are highly in accordance with the corresponding prior distributions. Indeed, we find that the posterior and prior distributions are almost indistinguishable, with only a couple of exceptions (one nuisance parameter in two out of the 20 redshift points) in which we find noticeable changes, which remain in any case small ($\lesssim1.9\sigma$). This allows us to sample the nuisance parameters of the GRBs at $z>2$ together with the nuisance and calibration parameters of those GRBs at $z<2$ to finally obtain the luminosity distances and build the extended distance ladder. In this way, we make sure to propagate the errors from the MCMC analysis to the final result. In particular, on top of these errors, we add the contribution of the intrinsic scatter to obtain the final uncertainties on the GRB luminosity distances.

In the upper plot of  Fig. \ref{fig:ext_DL}, we present the extended distance ladder obtained with our \textit{baseline} calibration without considering evolution effects and compare it with the one obtained when the latter are also taken into account. In particular, for the latter, we employ the results obtained when the evolutionary coefficients, $k_{L_{\rm X}}$ and $k_{L_{\rm peak}}$, are left free to vary according to the priors from the Whole sample since they lead to the tightest results and the smallest $\sigma_{int}$ for the Dainotti relations (see Sec. \ref{sec:evo}). To quantify possible deviations from the GP+CCH distances, for each calibration result, we compute the bias between their mean and the mean of the GP+CCH reconstruction at each redshift, $\beta = \overline{\log D_L^{\rm GRB}} - \overline{\log D_L^{\rm GP+CCH}}$. In the calibrated GRB data points of $\log D_L(z)$ obtained without evolution effects, displayed in the upper plot of Fig. \ref{fig:ext_DL}, one can already see a trend towards negative values of $\beta$ in the region $z>2$. The latter behavior was already found in previous works; see, e.g., \cite{postnikov14}. To keep track of this trend in a more precise way, we compute a binned bias. More concretely, we opt to bin the redshift range, composed of 50 points, in 10 equi-populated redshift bins. In order to account for the existing correlations between the various data points in each bin, we compute the weighted mean and its uncertainty as follows:

\begin{equation}\label{eq:bias}
    \tilde{\beta}_k = \frac{\sum_{i,j=e_{k}}^{f_{k}}\beta_{i}\omega_{ij}}{\sum_{i,j=e_{k}}^{f_{k}} \omega_{ij}},\\
    \tilde{\sigma}_k = \sqrt{\frac{1}{\sum_{i,j=e_{k}}^{f_{k}}\omega_{ij}}}\,,
\end{equation}%
with $\omega_{ij}=(C^{-1})_{ij}$ the elements of the inverse of the covariance matrix of the GRB luminosity distances, $e_{k}=1+5(k-1)$, and $f_k = 5k$. This is an accurate computation since the distribution of the GRB luminosity distances is Gaussian in very good approximation, as we have explicitly checked and in accordance with a recent work of \cite{Dainotti2023alternative}. We present our results in the middle and bottom panels of Fig. \ref{fig:ext_DL}, neglecting and considering the evolution corrections, respectively. In each of these two plots, we also show the weighted bias obtained when we neglect the off-diagonal terms of $C$. By comparing the middle and bottom plots, one can see that accounting for redshift evolution corrections is very important to decrease the bias in the region $z>2$. In fact, when no evolution effects are considered, there is a clear trend towards smaller luminosity distances, as already noted in the upper plot of Fig. \ref{fig:ext_DL}. We show now that correcting for evolutionary effects enables us to cure this trend, and this is even more true when we consider the existing correlations between the various GRB luminosity distances. Indeed, it is a well-known result that the physical properties of astrophysical objects at high $z$ are more affected by the Malmquist bias \citep{Malmquist1922} due to the difficulty of observing faint events at larger distances. This proves that if one wants to exploit probes at higher $z$ for cosmological purposes, it is of utmost importance to use correlations in which possible biases and effects of redshift evolution have been properly evaluated and corrected.

\section{Discussion}\label{sec:discussion}

All the analyses presented in Secs. \ref{sec:calib3D}-\ref{sec:calib2D} share a common result. Our method is stable under the choice of the priors on the parameters that characterize the plane $a,b,C_o,\sigma_{int}$. 
Even when using Gaussian priors at $1\sigma$, the posteriors on the parameters are compatible within $\sim 1-2\sigma$ to the other cases with wider priors.
The differences between all the cases analyzed are very small, although we achieve the tightest constraints for the 3D relation parameters when flat priors at 5$\sigma$ are employed (see for clarity Fig. \ref{fig:triangle_3D}). In addition, we find that the contribution of the peak luminosity in the prompt emission to the 3D Dainotti relation is not particularly favoured from a Bayesian perspective, but cannot be excluded either. Our results only set an upper bound on $b$. More specifically, we find that $b\lesssim 0.2$ at 95\% C.L. This result differs from what has been previously derived in analyses in which, to avoid the circularity problem, the parameters of the correlations are varied together with the cosmological parameters in a simultaneous fitting.
In that case, the 3D relation is strongly preferred over the 2D one. This preference is supported by the analysis in \citep{cao_gamma-ray_2022}, which considers six different flat and non-flat dark energy models and different GRBs data sets, which also include the full Platinum sample. However, our findings are not only fully independent of cosmological models but are also in agreement with theoretical predictions and reach a level of accuracy suitable for constraining cosmological parameters (see, in particular, the results in Sec. \ref{sec:calib2D}). Since we find $b$ to be compatible with 0 at 68\% C.L when we employ priors at 3 and 5$\sigma$, we feel motivated to also study the 2D relation in this work.

The baseline calibration both for the 2D and 3D Dainotti relations leads to an intrinsic scatter of the order of $\approx 0.2$ with a relative precision of $\sim20\%$. Specifically, for the 3D relation we obtain {$\sigma_{int} = 0.21^{+0.03}_{-0.05}$. If compared with standard analyses, such as the fundamental plane fitting result with the same number of GRBs (see Table \ref{tab:prior_3D}), we obtain a decrease of $\sim40\%$ in its central value, outlining the achievement of a tighter relation.
In particular, we can also quantify the compatibility of the two aforementioned results by calculating the relative difference of $\sigma_{int,\rm FP}$ compared to $\sigma_{int}$ normalized by the maximum of the two uncertainties which reads as: $(\sigma_{int,\rm FP}- \sigma_{int})/ \delta$, with $\delta=\max(\delta_{\sigma_{int}},\delta_{\sigma_{int,\rm FP}})$. We find it to be 1.99, showing that the calibration result is consistent within $2\sigma$ with the standard (model-dependent) analysis results. As a term of comparison, employing a different OHD data set and thus a different set of 23 GRBs, \cite{Tian:2023ffe} calibrate the 3D Dainotti relation with the D'Agostini method and obtain a very large scatter, $\sigma_{int}=0.70\pm0.10$.
In general, a lower dispersion is expected when GRBs are employed closer to the plane. One example can be found in \citep{Dainotti22}, where the 10 GRBs closest to the plane are selected from the Platinum sample, leading to $\sigma_{int} = 0.05\pm0.05$ (assuming a flat $\Lambda$CDM model). This analysis has the scope to show how many of these GRBs should be used in the future once more data is available in order to achieve the same precision of the SNIa standard candles in constraining cosmological parameters. In this regard, we can ask ourselves what impact the intrinsic scatter of the GRBs has on the calibration with CCH. At the end of Sec. \ref{sec:calib3D}, we actually perform an analysis to answer this question. Indeed, we have seen that our results remain unchanged whether or not we include the uncertainties on the reconstruction of the luminosity distance in the likelihood. This result holds for both the 2D and 3D relations and has two main implications. 
First, it confirms even more the stability of all the results obtained throughout this work. It establishes the CCH as a reliable calibrator for pinpointing a specific GRB sub-set with low intrinsic scatter. This is remarkable if one considers that this result holds even when the redshift evolution corrections are accounted for, thus making the relation and the results themselves more robust if also compared with previous results in the literature. Indeed, we obtain a mode value for $\sigma_{int}$ which stands between 0.18 and 0.20, the lowest being obtained when also the evolutionary coefficients are left free in the analysis. As a comparison, \cite{dainotti2022gamma} find $\sigma_{int}=0.22\pm0.10$, and most recently this value has been better constrained in \citep{dainotti2023gamma}, where they find $\sigma_{int} = 0.18\pm0.07$. Both results are obtained using the full Platinum sample and assuming a flat $\Lambda$CDM model. When we compare the uncertainty on this latter result including the correction for evolution, $\delta_{\sigma_{int}}=0.07$, with our result, $\delta_{\sigma_{int}}=0.04$, we obtain a decrease of 43\%. 
This result makes this method competitive with standard approaches, especially for using GRBs for cosmological purposes in light of the already discussed circularity problem. Additionally, there is still a non-negligible dispersion on the GRB plane compared to other probes. As it stands, this dispersion is sufficiently large to prevent a better calibration regardless of the potential increase in the amount of CCH data at $z\lesssim2$ through dedicated galaxy surveys in the future (see, e.g., \cite{moresco2022unveiling}).

Fig. \ref{fig:ext_DL} is exactly telling us that, although GRB data alone cannot achieve yet the same constraining power as other low-redshift probes (see the difference in the error bars with respect to GP+CCH at the lowest redshifts), their strength lies precisely in their capability of extending the ladder to higher redshifts.
We remark here that this has motivated in the last two decades the effort to build several correlations that can be used in this context, provided that understanding of their physical meaning and reliability regardless of selection biases do hold. However, these two aspects have not yet been fully addressed in many of the correlations. These correlations have been similarly calibrated with low-$z$ anchors, and their intrinsic scatter ranges from $\sim$ 0.2 to 0.55.
For instance, the Amati relation has been widely calibrated with SNIa using GP or simultaneous fitting analyses. The former approach was followed, for example, by \cite{liu_gamma-ray_2022} and \cite{liang_calibrating_2022}, who found an intrinsic scatter between $0.46$ and $0.52$. The last authors further conducted simultaneous fitting with 31 OHD within the $\Lambda$CDM and $w$CDM models, yielding central values ranging from $0.39$ to $0.46$, depending on whether the A220 or A118 sample is used \citep{Khadka:2021vqa}. \cite{kumar_gamma_2023} calibrated the Amati relation using GP and 32 CCH, along with the A220. They determined the scatter to be $0.289^{+0.015}_{-0.014}$. To mention a few others, \cite{li_testing_2023} used SNIa with GP in the A118 GRB sample at redshift less than $1.4$, finding central values of $0.49$ to $0.55$, depending on the likelihood employed, whereas \cite{Mu:2023bsf} obtained a dispersion of $0.54$ with SNIa in conjunction with the A220 sample. \cite{Alfano:2024ukk} found a scatter around $0.36$ employing a simultaneous fitting for the Yonetoku relation. In light of these mentioned results, it is evident that a common low dispersion is not still reached for relations based solely on the prompt emission phase. The result achieved in this work represents so far, and at the best of our knowledge, one of the tightest correlations present in the literature. One possible explanation could be related to the fact that the plateau phase is less varied in its intrinsic properties with respect to the prompt emission. The scatter we find is robust to corrections for selection biases and redshift evolution, in contrast with other correlations, in which these effects still play a non-negligible role. If these corrections significantly affect the correlations, the correlations will no longer exist intrinsically in a reliable way. Also, the physical explanation of a relation meant to be used as a standard candle should be clear. For the correlations analyzed in this work, the most accredited theoretical scenario is the magnetar emission, and we actually found our results to support this model as the slope of the correlation is compatible with -1. These aspects make the extended ladder, which can be regarded as the final product of this calibration, even more robust and powerful to be employed in future cosmological applications.

\section{Summary and Conclusions}\label{sec:conclusion}

In this work, we have calibrated the Dainotti relations through a model-independent method that makes use of cosmic chronometers as calibrators. Thanks to the stability of our results across all the analyses presented, we can conclude that these low-redshift data are capable of identifying a valuable set of standardizable candles}. Indeed, through the CCH calibration, the 20 GRBs we analyzed in $0.553\leq z \leq 1.96$ are found to adhere tightly to the fundamental plane. In particular, we achieved one of the lowest levels of intrinsic scatter observed to date, and we derived constraints on the 2D and 3D Dainotti relations.
These constraints are compatible with the physics of the relation itself, thus supporting, even more, its theoretical interpretation. This set can, therefore, be a promising candidate for future use in cosmological applications, as we have already demonstrated in this work by extending the distance ladder up to $z=5$ through an unbiased GRB correlation.

Finding novel distance indicators that can be both less affected by biases and systematics and capable of extending the range of applicability of the cosmic distance ladder method to larger redshifts has become an important challenge in cosmology and astrophysics. However, if we aim to obtain unbiased cosmological distances, the calibration procedure cannot be subject to strong model-dependent assumptions. This is particularly relevant in light of the existing cosmological tensions, such as the one on $H_0$. This prospect further motivates ongoing efforts to achieve increasingly precise measurements, enhancing the statistics at low redshifts and in regions where objects like GRBs or QSOs can be crucial. On the other hand, the spread in the GRBs observed luminosities is still a non-negligible issue due to the not yet well-defined nature of their origin (core collapse of a massive star, merger of two neutron stars in a binary system or a neutron star-black hole system merger). Therefore, it becomes essential to investigate the reliability of these alternative probes and the relations that govern their intrinsic properties, possibly taking advantage also of the improvement both in data quality and quantity with upcoming surveys, e.g., SVOM \citep{atteia2022svom} or THESEUS \citep{amati2021theseus} missions. In this way, model-independent calibration methods could play a significant role in standardizing these probes effectively, thereby further extending the cosmic distance ladder and broadening our ability to probe the fundamental cosmological parameters.

\appendix

\section{Data sets}\label{app:data}
\setcounter{table}{0}

In this appendix, we present the tables with the GRBs and the CCH data employed in this work and described in Secs. \ref{sec:platinum} and \ref{sec:cch}, respectively.

\begin{table*}[ht!]
    \centering                                     
\begin{tabular}{c c c c c c c c c c c c}          
\hline\noalign{\smallskip}  
ID & $z$ & $\log T^*_X$ & $\sigma_{\log T^*_X}$ & $F_{\rm peak}$ & $\sigma_{F_{\rm peak}}$ & $K_{\rm prompt}$ & $\sigma_{K_{\rm prompt}}$ & $\log F_X$ & $\sigma_{\log F_X}$ & $K_{\rm plateau}$ & $\sigma_{K_{\rm plateau}}$ \\
\hline\noalign{\smallskip}                             
        070521  & 0.553 & 3.361 & 0.048 & 8.71e-08 & 1.20e-08 & 0.662 & 0.053 & -10.012 & 0.049 & 0.991 & 0.175 \\
 080430   & 0.767 & 3.970 & 0.038 & 1.82e-07 & 8.08e-09 & 0.868 & 0.036 & -11.172 & 0.028 &   3.459 & 2.059 \\
 151027A & 0.810 & 3.741 & 0.017 & 5.82e-07 & 2.88e-08 & 0.637 & 0.031  &  -9.935 & 0.022 &  0.956 & 0.021 \\
 070508   & 0.820 & 2.765 & 0.012 & 2.24e-06 & 3.07e-08 & 0.570 & 0.032 &  -9.161 & 0.011 &  0.868 & 0.012 \\
 060814   & 0.840 & 3.959 & 0.035 & 6.06e-07 & 1.48e-08 & 0.653 & 0.016 & -10.911 & 0.035 &  0.982 & 0.036 \\
 140506A & 0.889 & 3.031 & 0.069 & 8.67e-07 & 4.56e-08 &  0.664 & 0.105 &  -9.904 & 0.055 &  0.938 & 0.136 \\
 091018   & 0.971 & 2.602 & 0.034 & 5.91e-07 & 1.22e-08 & 1.140 & 0.030 &  -9.627 & 0.024 &  0.941 & 0.041 \\
 131030A & 1.290 & 2.486 & 0.028 & 2.65e-06 & 4.61e-08 & 0.416 & 0.041 &  -9.294 & 0.025 &  0.775 & 0.017 \\
 061121   & 1.314 & 3.423 & 0.017 & 1.96e-06 & 2.43e-08 & 0.451 & 0.008 & -10.030 & 0.013 &  0.958 & 0.037 \\
 150910A & 1.360 & 3.481 & 0.029 & 8.28e-08 & 2.18e-08 & 0.632 & 0.309 &  -9.971 & 0.704 &  0.750 & 0.037 \\
 110213A & 1.460 & 3.450 & 0.024 & 7.15e-08 & 1.88e-08 & 2.100 & 0.903 &  -9.897 & 0.036 &  1.187 & 0.295 \\
 060418   & 1.490 & 2.723 & 0.057 & 4.99e-07 & 1.63e-08 & 0.633 & 0.031 &  -9.793 & 0.049 &  0.982 & 0.122 \\
 070306   & 1.496 & 4.472 & 0.028 & 2.01e-07 & 8.99e-09 & 0.639 & 0.408 & -11.274 & 0.034 &  0.857 & 0.040 \\
 090418A & 1.608 & 3.112 & 0.031 & 1.58e-07 & 1.66e-08 & 0.508 & 0.082 & -10.047 & 0.029 &  0.981 & 0.083 \\
 131105A & 1.686 & 3.520 & 0.048 & 2.68e-07 & 2.04e-09 & 0.582 & 0.064 & -10.911 & 0.033 &  0.942 & 0.093 \\
 091020   & 1.710 & 2.520 & 0.045 & 3.64e-07 & 1.71e-08 & 0.443 & 0.033 &  -9.713 & 0.034 &  0.901 & 0.067 \\
 150314A & 1.758 & 2.052 & 0.016 & 3.81e-06 & 6.06e-08 & 0.297 & 0.033 &  -8.610 & 0.015 &  0.760 & 0.015 \\
 190106A & 1.860 & 3.735 & 0.031 & 3.36e-07 & 1.30e-08 &  0.614 & 0.073 & -10.432 & 0.029 &  2.675 & 2.089 \\
 060708   & 1.920 & 3.040 & 0.063 & 6.89e-08 & 7.96e-09 & 0.467 & 0.563 & -10.869 & 0.061 &   1.682 & 0.803 \\
 160121A & 1.960 & 3.353 & 0.120 & 7.79e-08 & 8.23e-09 & 0.822 & 0.136 & -11.173 & 0.069 &   1.022 & 0.161 \\
\hline                                            
\end{tabular}
\caption{Sub-sample of the 20 GRBs from the Platinum data set \citep{dainotti2020x} used in the calibration of the 3D and 2D Dainotti relations. $T^*_X$ has units of $s$, while the fluxes $F_{\rm peak}$ and $F_{\rm X}$ are given in $erg \ cm^{-2} \ s^{-1}$. The $K$-corrections are dimensionless. For all the quantities we also report the associated 1$\sigma$ error. All these objects are provided by the \textit{third Swift GRB Catalog} \citep{lien2016third}, except for the 091018 and 131105A GRBs, which are from the \textit{Swift BAT burst analyser} \citep{evans2010swift}. We refer the reader to Table A1 of \cite{cao_gamma-ray_2022} for the remaining GRBs of the Platinum sample.}
    \label{tab:20platinum}
\end{table*}%

\begin{table*}[ht!]
\centering                                     
\begin{tabular}{c c c c c c}          
\hline\noalign{\smallskip}  
$z$ & $H(z)$ \ [Km/s/Mpc] & References & $z$ &$ H(z) $\ [Km/s/Mpc] & References \\   
\hline\noalign{\smallskip}                             
   0.07 & $69.0\pm19.6$ & \text{\citep{Zhang:2012mp}} &  0.48 & $97.0\pm62.0$ & \text{\citep{Stern:2009ep}} \\ 
        0.09 &$ 69.0\pm12.0$ & \text{\citep{Jimenez:2003iv}} & 0.5929 & $107.0\pm15.5$ & \text{\citep{moresco2012improved}} \\
        0.12 & $68.6\pm26.2$ & \text{\citep{Zhang:2012mp}} & 0.6797 & $95.0\pm10.5 $& \text{\citep{moresco2012improved}} \\
        0.17 & $83.0\pm8.0 $& \text{\citep{Simon:2004tf}} &  0.75 & $98.8\pm33.6$ & \text{\citep{Borghi:2021rft}} \\
        0.1791 & $78.0\pm6.2$ & \text{\citep{moresco2012improved}} &   0.7812 &$ 96.5\pm12.5$ & \text{\citep{moresco2012improved}} \\
        0.1993 & $78.0\pm6.9$ & \text{\citep{moresco2012improved}} &  0.8754 & $124.5\pm17.4$ & \text{\citep{moresco2012improved}} \\
        0.2 & $72.9\pm29.6$ & \text{\citep{Zhang:2012mp}} &  0.88 &$ 90.0\pm40.0 $& \text{\citep{Stern:2009ep}}\\
        0.27 & $77.0\pm14.0$ & \text{\citep{Simon:2004tf}}  & 0.9 &$ 117.0\pm23.0$ & \text{\citep{Simon:2004tf}}\\
        0.28 &$ 88.8\pm36.6$ & \text{\citep{Zhang:2012mp}} & 1.037 &$ 133.5\pm17.6$ & \text{\citep{moresco2012improved}}\\
        0.3519 & $85.5\pm15.7 $& \text{\citep{moresco2012improved}} &  1.26 &$ 135.0\pm65.0$ & \text{\citep{Tomasetti:2023kek}}\\
        0.3802 & $86.2\pm14.6 $& \text{\citep{Moresco:2016mzx}} & 1.3 &$ 168.0\pm17.0 $& \text{\citep{Simon:2004tf}}  \\
        0.4 & $95.0\pm17.0$ & \text{\citep{Simon:2004tf}}  & 1.363 & $160.0\pm33.8 $& \text{\citep{Moresco:2015cya}} \\
        0.4004 & $79.9\pm11.4$ & \text{\citep{Moresco:2016mzx}} & 1.43 & $177.0\pm18.0 $& \text{\citep{Simon:2004tf}}  \\
        0.4247 & $90.4\pm12.8 $& \text{\citep{Moresco:2016mzx}} &  1.53 & $140.0\pm14.0$ & \text{\citep{Simon:2004tf}}  \\
        0.4497 & $96.3\pm14.4$ & \text{\citep{Moresco:2016mzx}} & 1.75 &$ 202.0\pm40.0$ & \text{\citep{Simon:2004tf}}  \\
        0.47 & $89.0\pm49.6$ & \text{\citep{Ratsimbazafy:2017vga}} & 1.965 & $186.5\pm50.6 $& \text{\citep{Moresco:2015cya}} \\
        0.4783 &$83.8\pm10.2 $& \text{\citep{Moresco:2016mzx}} \\       
\hline                                            
\end{tabular}
\caption{In this table, we present the 33 CCH data points used in this work, with the corresponding references. All the uncertainties are evaluated at 68\% C.L. and, in the case of Refs. \cite{moresco2012improved, Moresco:2016mzx}, the central values of $H(z)$ are computed by performing the arithmetic mean of the measurements obtained with the BC03 \citep{Bruzual:2003tq} and MaStro \citep{Maraston:2011sq} SPS models. We remind the reader that we incorporate in the analysis also the correlations through the covariance matrix from \cite{Moresco:2020fbm}.}\label{tab:CCH}
\end{table*}%

\section{Kolmogorov-Smirnov and Anderson-Darling tests for the Platinum sample}\label{app:KSAD}
\setcounter{table}{0}
In this appendix, we show the results of the KS and AD tests performed to ensure that the sub-sample of 20 GRBs used to calibrate the Dainotti relations with CCH at $z\lesssim2$ are fairly representative of the entire population of GRBs (see Sec. \ref{sec:platinum}). This is our null hypothesis. We apply the KS and AD tests to each variable of the Platinum sample employed in this work: $\log T^{\star}_X$, $F_{\rm peak}$, $K_{\rm prompt}$, $\log F_X$ and $K_{\rm plateau}$. We decide to not reject the null hypothesis if the $p$-value $p\geq0.05$ and reject it if $p<0.05$. 
The results of both the KS and the AD test are displayed in Table \ref{tab:KSAD}. The $p$-values listed in the table show that we cannot reject the null hypothesis in all cases, regardless of the variable in the Platinum sample employed in the test.

\begin{table}[t!]
    \centering
    \begin{tabular}{cccccc}
    \hline
         & $\log T_X^{\star}$ &  $F_{\rm peak}$ & $K_{\rm prompt}$ & $\log F_X$ & $K_{\rm plateau}$   \\
         \hline
         $p$-value (KS) & 0.949 & 0.504 & 0.254 & 0.391 &  0.074 \\
         $p$-value (AD) &  0.250 &  0.250& 0.116 & 0.250 & 0.052 \\
         \hline
    \end{tabular}
    \caption{$p$-values obtained from the KS and the AD tests performed to test if the sub-sample of 20 GRBs at $z\lesssim2$ used to calibrate the Dainotti relations with CCH and the 50 GRBs from the full Platinum sample are likely to be obtained from the same parent distribution. This is our null hypothesis. All the $p$-values displayed in the table allow us not to reject the null hypothesis, since $p\geq0.05$.}
    \label{tab:KSAD}
\end{table}


\section*{Acknowledgements}

AF and MM acknowledge support from the INFN project “InDark”. AGV is funded by “la Caixa” Foundation (ID 100010434) and the European Union's Horizon 2020 research and innovation programme under the Marie Sklodowska-Curie grant agreement No 847648, with fellowship code LCF/\\BQ/PI21/11830027. MM is also supported by the ASI/LiteBIRD grant n. 2020-9-HH.0 and by the Italian Research Center on High Performance Computing Big Data and Quantum Computing (ICSC), project funded by European Union - NextGenerationEU - and National Recovery and Resilience Plan (NRRP) - Mission 4 Component 2 within the activities of Spoke 3
(Astrophysics and Cosmos Observations). AF, MGD and AGV also acknowledge the participation in the COST Action CA21136 “Addressing observational tensions in cosmology with systematics and fundamental physics” (CosmoVerse).

\bibliographystyle{elsarticle-harv}
\bibliography{main}






\end{document}